\documentclass[letter,prd,aps,floatfix,superscriptaddress,nofootinbib]{revtex4-1}
\usepackage{amssymb,mathrsfs}
\usepackage{graphicx,float,amsmath,amssymb}
\usepackage{hyperref}
\usepackage[usenames,dvipsnames]{color}
\usepackage{array}

\begin{document}


\title{Challenges to global solutions in Horndeski's theory}

\author{Laura Bernard} 
\affiliation{Perimeter Institute for Theoretical Physics, Waterloo, ON N2L 2Y5, Canada}
\affiliation{LUTH, Observatoire de Paris, PSL Research University, CNRS, Universit\'e Paris Diderot, Sorbonne Paris Cit\'e, 5 place Jules Janssen, 92195 Meudon, France}
\author{Luis Lehner}
\affiliation{Perimeter Institute for Theoretical Physics, Waterloo, ON N2L 2Y5, Canada}
\author{Raimon Luna}
\affiliation{Perimeter Institute for Theoretical Physics, Waterloo, ON N2L 2Y5, Canada}
\affiliation{Departament de F\'{\i}sica Qu\`antica i Astrof\'{\i}sica, Institut
  de Ci\`encies del Cosmos, Universitat de Barcelona, Mart\'{\i} i Franqu\`es 1,
  E-08028 Barcelona, Spain}
\date{\today}

\def\lgrho{\log_{10}\rho}

\begin{abstract}
We explore the question of obtaining global solutions in Horndeski's theories
of gravity. Towards this end, we study a relevant set of the theory and, by
employing the Einstein frame we simplify the analysis by exploiting known results 
on global solutions of wave equations.
We identify conditions for achieving global solutions as well as obstacles that 
can arise to spoil such goal. We illustrate such problems via numerical simulations.
\end{abstract}

\maketitle

\section{Introduction}
In the context of extensions to General Relativity, Horndeski's theories~\cite{1974IJTP...10..363H} stand
out since they constitute the most general four-dimensional, diffeomorphism
covariant theories leading to second order equations of motion~\cite{1974IJTP...10..363H,Woodard:2006nt,Deffayet:2011gz}.
These theories, where the gravitational degrees of freedom are expressed
in terms of a metric tensor together with a scalar field, have a particularly distinct feature.
Namely, the equations of motion they define are of {\em second order type}.
This property  ensures the absence of Ostrogradski  ghosts (e.g.~\cite{Woodard:2006nt}), which is arguably a necessary condition for any physical theory. 
The freedom allowed by this family of 
theories  has been exploited in many fronts, for instance, with goals to address dark energy (see e.g.~\cite{Nicolis:2008in,Deffayet:2009wt,Ferreira:2019xrr,Kobayashi:2011nu}
and references therein) and
incipient explorations of  extensions of GR in binary mergers (e.g.~\cite{Damour:1996ke,Barausse:2012da,Mirshekari:2013vb,Sagunski:2017nzb,Julie:2017pkb,Hirschmann:2017psw,Bernard:2018hta,Witek:2018dmd}).
In the former front, particular examples include quintessence \cite{Ratra:1987rm, Caldwell:1997ii}; kinetic quintessence or k-essence 
\citep{ArmendarizPicon:2000dh,ArmendarizPicon:2000ah} and chameleon/galileon \citep{Khoury:2003aq,Khoury:2003rn}
theories. 

Typically, applications have been analyzed in the context of linearized studies over specific
backgrounds endowed with special symmetries, and consistency with observations in different regimes is
addressed. In particular, the subset of Horndeski's theories allowed by observation have been severely constrained by the gravitational wave event GW170817~\cite{Baker:2017hug,Creminelli:2017sry,Ezquiaga:2017ekz},
with constraints derived from linearized studies. 
In spite of certainly relevant discussions and consequences derived in this context, 
it is important however to understand the fully nonlinear behavior of allowed theories
within this family.  For instance, a successful observation
derived at the linear level which is not derivable from the nonlinear system would arguably call into question
the original action as providing a true model for nature\footnote{In the former case, the linear system should instead be 
regarded as the fundamental building block} or the extent to which a linear analysis
yields comprehensive/sensible constraints. Consistency of solutions to the linear problem with those
of the full problem in the linear regime is not always a given as non-linear solutions might give rise to
phenomenology completely absent at the linear level --even when initial conditions are chosen within
the linear regime.  In the case of General Relativity and its application in cosmology, 
requirements for this being the case have been discussed in~\cite{fischer1973}.

Naturally, the extent to which a linear solution can be trusted --with respect to the
original action-- depends 
on one's understanding of the non-linear regime. To gain such understanding, which in turn can help
decide which subset within the theory can be considered physical, a stricter and certainly natural
condition must be satisfied. Namely,  {\em well posedness}~\cite{jH02}. This condition, implies a given
problem has a unique solution that depends continuously on initial and boundary data\footnote{The absence
of Ostrogradski ghosts is necessary but certainly not sufficient for well posedness.}. From a natural
point of view, the satisfaction of these properties is crucial for the physical understanding of the problem 
under consideration.

Consequently, insisting a valid theory must yield well-posed initial boundary value problems
is a general powerful requirement to restrict potential options and identify possible alternatives to
General Relativity. The analysis of this condition in specific theories is typically complex, 
which has hindered drawing straightforward conclusions for Horndeski's theories in the past. 
Recent works however have begun to explore this issue and, in particular, have uncovered
significant restrictions~\cite{Papallo:2017qvl,Ijjas:2018cdm,Kovacs:2019jqj} even locally. 
We stress the importance of this condition can not 
be underestimated even in linearized regimes. Failure to satisfy 
it implies it would be impossible to trust possible solutions let alone seek them in the first place
--regardless of the method employed to seek for such solutions!

In this note, both strongly motivated by and building from these recent works, we revisit the problem 
from a slightly
different angle by performing our analysis in a different frame --the so called
Einstein frame--. As a result, a simpler system is obtained where the complexity of the
analysis lies primarily in understanding the equation of motion for the scalar field.
In such frame it is arguably easier to elucidate the degree to which these theories display a more 
involved behavior when contrasted with General Relativity and the obstacles that might arise to
obtain global solutions. 

We recall that much has been discussed on the issue of frames in scalar-tensor gravity theories
such as Horndeski's (e.g.~\cite{Faraoni:1999hp,Flanagan:2004bz}).  Whether
the Ricci scalar appears ``clean'' in such action or not is often
signaled as being a description in the Einstein (former) or Jordan frame.
Here we explore (a subset of) Horndeski's theories from the initial value problem (IVP) point 
of view in the Einstein frame,  discuss the existence of delicate issues and illustrate their 
consequences via numerical simulations. As well, and for concreteness, we do not concern ourselves
at this time with the ``non-vacuum case'' --i.e. scenarios which include matter. Here further issues
come into play that in and on themselves raise further concerns for the theory to remain viable
--from observational points of view. This, in turn, requires invoking mechanisms like ``Vainshtein
screening'' which bring about further mathematical difficulties, see e.g.~\cite{Brito:2014ifa}.

This work is organized as follows. In section II we revisit the special 
case of Horndeski's theories that has been identified as well posed and re-analyze
it in the Einstein frame drawing general conclusions about such case and discuss
issues related to well posedness in the nonlinear regime. In section III and IV we illustrate our 
discussion both analytically and numerically in a couple of special cases. 


\section{Horndeski's theory, special case analyzed}
Horndeski's theories describe gravitational interactions in terms of a metric
tensor $g_{ab}$ and a scalar field $\phi$. Their equations of motion
are determined from the action,
\begin{equation}
S = \frac{1}{16\pi G} \int d^4x \sqrt{-g} (\Sigma_{i=1}^{5} \, {\cal L}_i)
\end{equation}
where,
\begin{eqnarray}
{\cal L}_1 &=& R + X - V(\phi) \, , \\
{\cal L}_2 &=& {\cal G}_2(\phi,X) \, , \\
{\cal L}_3 &=& {\cal G}_3(\phi,X) \Box \phi  \, , \\
{\cal L}_4 &=& {\cal G}_4(\phi,X) R + \partial_X {\cal G}_4(\phi,X) \delta^{ac}_{bd} \nabla_a \nabla^b \phi \nabla_c \nabla^d \phi  \, , \\
{\cal L}_5 &=& {\cal G}_5(\phi,X) G_{ab} \nabla^a \nabla^b \phi - \frac{1}{6} \partial_X {\cal G}_5(\phi,X) \delta^{ace}_{bdf}
 \nabla_a \nabla^b \phi \nabla_c \nabla^d \phi  \nabla_e \nabla^g \phi  \, .
\end{eqnarray}
with $X=-1/2 \nabla_a \phi \nabla^a \phi$, $G_{ab}$ the Einstein tensor, ${\cal G}_i$ are functions of the
scalars $\{ \phi,X \}$, $V$ is a potential and $\delta_{a_1..a_n}^{b_1..b_n}$ is the generalized Kronecker delta symbol.

A thorough analysis of hyperbolicity properties of the resulting equations of motion, given the complexity of the PDE system,
is naturally a difficult task. One such study has been presented recently in~\cite{Papallo:2017qvl} (see
also~\cite{Ijjas:2018cdm}). Here, following steps taken to establish local well posedness of Einstein equations~\cite{fourès-bruhat1952},
--whereby the introduction of harmonic coordinates renders Einstein equations manifestly symmetric hyperbolic-- a judicious
coordinate choice is found to guarantee strong hyperbolicity. Within this context, it is shown that only a special subset 
of Horndeski's theories leads to strong hyperbolicity in harmonic gauge in the nonlinear regime. 
This subset is given by the action,

\begin{equation}\label{jordanH1}
S=\frac{1}{16\pi}\int d^4x \sqrt{-g} \left[ 
(1+{\cal G}_4(\phi))R 
+ X - V(\phi)  
+ {\cal G}_2(\phi,X) \right] \; , 
\end{equation}  
Notice the action above corresponds to the so called
Jordan frame (as the Ricci scalar appears multiplied by a non-trivial function of $\phi$). The equations of
motion obtained from this action can be found in~\cite{Papallo:2017qvl}. A conformal
transformation, of the form $\tilde g_{ab} = \Omega^2 g_{ab}$ with $\Omega=\sqrt{1+{\cal G}_4(\phi))}$, can be exploited to obtain the equations in the Einstein frame. We assume that the conformal factor $\Omega$ never vanishes, which ensures that the transformation is well-defined and the two formulations of the theory are equivalent. It allows one to rewrite the above action as,

\begin{equation}\label{einsteinH1}
S=\frac{1}{16\pi}\int d^4x \sqrt{-\tilde g} \left\{ 
\tilde R 
+ \frac{1}{(1+{\cal G}_4(\phi))^2}\left[\left(3[{\cal G}'_4(\phi)]^2 +1+{\cal G}_4(\phi)\right)\tilde{X} - V(\phi)  
+ {\cal G}_2\left(\phi,(1+{\cal G}_4(\phi))\tilde{X}\right)\right] \right\} \; , 
\end{equation} 
where $\tilde X=-1/2 \tilde{\nabla}_c \phi \, \tilde{\nabla}^c \phi$. 
From this action, the equations of motion are
\begin{align}
& \tilde{G}_{ab} = \left[\frac{3[{\cal G}'_4(\phi)]^2+1+{\cal G}_4(\phi)}{2(1+{\cal G}_4(\phi))^2}\,\tilde{X} +\frac{- V(\phi) + {\cal G}_2(\phi,X)}{2(1+{\cal G}_4(\phi))^2}\right]\tilde{g}_{ab} +\left[\frac{3[{\cal G}'_4(\phi)]^2}{2(1+{\cal G}_4(\phi))^2}+\frac{1+\partial_{X}{\cal G}_{2}(\phi,X)}{2(1+{\cal G}_4(\phi))}\right]\tilde{\nabla}_a \phi\tilde{\nabla}_b \phi \;,\label{einsteingabpregauge} \\[7pt] \nonumber
& \biggl[\tilde{g}^{ab} - \frac{(1+{\cal G}_{4}(\phi))^2 \partial^2_{XX}{\cal G}_{2}(\phi,X)}{3[{\cal G}'_4(\phi)]^2+(1+{\cal G}_4(\phi))(1+\partial_{X}{\cal G}_{2}(\phi,X))} \tilde{\nabla}^a \phi\tilde{\nabla}^b \phi \biggr] \tilde{\nabla}_a\tilde{\nabla}_b \phi  \\ \nonumber
&\qquad\qquad = \frac{1}{3[{\cal G}'_4(\phi)]^2+(1+{\cal G}_4(\phi))(1+\partial_{X}{\cal G}_{2}(\phi,X))} \Biggl\{V'(\phi)-\partial_{\phi}{\cal G}_{2}(\phi,X)-2\,{\cal G}'_4(\phi)\frac{V(\phi)-{\cal G}_{2}(\phi,X)}{1+{\cal G}_{4}(\phi)} \\ \nonumber
&\qquad\qquad\qquad +\biggl[2(1+{\cal G}_{4}(\phi))\partial^{2}_{\phi X}{\cal G}_{2}(\phi,X)+{\cal G}'_{4}(\phi) \Bigl( 6{\cal G}_4''(\phi)-1-3\partial_{X}{\cal G}_{2}(\phi,X)-6\frac{[{\cal G}'_4(\phi)]^2}{1+{\cal G}_{4}(\phi)}\Bigr)\biggr]\tilde{X} \\
&\qquad\qquad\qquad +2{\cal G}'_{4}(\phi)(1+{\cal G}_{4}(\phi))\partial^{2}_{XX}{\cal G}_{2}(\phi,X)\tilde{X}^{2} \Biggr\} \label{einsteinphipregauge} \;.
\end{align}
In order to write the scalar field equation in the convenient form~\eqref{einsteinphipregauge}, we have divided it by the overall factor ${{3[{\cal G}'_4(\phi)]^2+(1+{\cal G}_4(\phi))(1+\partial_{X}{\cal G}_{2}(\phi,X))}\,(1+{\cal G}_{4}(\phi))^{-2}}$. In the following, we will assume that this factor is non-zero.
Notice that neither of the right hand sides in eqns~\eqref{einsteingabpregauge}-\eqref{einsteinphipregauge} involve second order derivatives of the relevant fields (metric or scalar),
and, the  hyperbolic properties of the system can be assessed independently for $g_{ab}$ and $\phi$.
In the case of the metric tensor, such properties only depend on the metric tensor itself and we can draw from the vast
knowledge about properties of Einstein equations (see e.g.~\cite{Sarbach:2012pr}).  In particular, we recall that
they can be straightforwardly rendered into symmetric hyperbolic form.
Indeed, following again~\cite{fourès-bruhat1952}, one can introduce harmonic 
coordinates ($\tilde \Gamma_a = 0$), and equation~\eqref{einsteingabpregauge} becomes
symmetric hyperbolic. Further, we recall the speed of propagation
of perturbations is {\em independent of the metric tensor itself} (thus the equation is linearly degenerate 
and no shocks can arise from smooth initial data). Importantly, the observations above with regards to well posedness
(at least locally) and linear degeneracy are certainly valid for other gauges. As we shall discuss below, regardless
of the gauge choice, the scalar field equation has particular `worrisome' properties.

The principal part of the scalar field equation depends on $\{g_{ab},\phi,\partial_a \phi\}$. Indeed,
the principal part of the equation for $\phi$, equation (\ref{einsteinphipregauge}), is given by a wave equation of a modified metric
\begin{equation}
\gamma^{ab} = \tilde{g}^{ab}  - 
   \frac{(1+{\cal G}_{4}(\phi))^2 \partial^2_{XX}{\cal G}_{2}(\phi,X)}{3[{\cal G}'_4(\phi)]^2+(1+{\cal G}_4(\phi))(1+\partial_{X}{\cal G}_{2}(\phi,X))} \tilde{\nabla}^a \phi\tilde{\nabla}^b \phi \label{effectiveinversemetric}\, .
\end{equation}
Thus, propagation speeds of scalar field perturbations depend on the state of the field and its gradient.
As a consequence, shocks can develop from smooth initial data, at which point uniqueness of the solution is lost and with it, 
well posedness~\footnote{To recover it, further conditions would need to be imposed, see discussions 
in~\cite{Reall:2014sla,Allwright:2018rut}.}.
Another potential problem is that the equation itself might change character
point-wise in the spacetime. Indeed the character of this equation, i.e. hyperbolic, elliptic or
parabolic, is determined by the eigenvalues of $\gamma^{ab}$. Namely, if no eigenvalue is zero, and the sign
of only one of them is opposite to the others the equation is hyperbolic~\footnote{If more than one is
of opposite sign, the equation would be ultra-hyperbolic in character.} (with $+$ signature it would be one negative).
If all signs are the same the equation is elliptic and if at least one eigenvalue is zero parabolic.
For a well defined initial value problem describing a small departure from General Relativity, the equation would
be hyperbolic. Notice that at the linear level, equation~\eqref{einsteinphipregauge} is symmetric hyperbolic, linearly degenerate and 
the  scalar field  perturbations propagate at the speed of light of the metric $\tilde g_{ab}$.
However, at the non-linear level --even with smooth initial data-- if dispersion does not win and gradients 
grow (assuming $\partial_{XX}{\cal G}_2 \neq 0$) the character of the equation 
can change and,  by continuity, it would do so by turning --locally-- to parabolic and then elliptic. Thus 
either through a change of character, or by loss of uniqueness due to shocks well posedness could be lost.

Interestingly, a change in character in  spherically
symmetric non-linear studies in subclasses of Horndeski's theories has been identified, for instance in 
k-essence~\cite{Akhoury:2011hr} and  Einstein-Dilaton-Gauss-Bonnet~\cite{Ripley:2019hxt}. The theories
studied in these references are seemingly different from Eq.~\eqref{jordanH1}, but they can be linked to a Horndeski theory through the following mappings. In the former case, only the kinetic term ${\cal G}_{2}(X)$ is present, while in the latter we only have ${\cal{G}}_{5}(\phi,X)=-\lambda\ln|X|$ where $\lambda$ is the coupling constant~\footnote{Although such a function ${\cal G}_{5}$ is not smooth at $X=0$, the equations of motion are well defined everywhere~\cite{Papallo:2017ddx}.}.

Notice however that the potential change in character or the development of shocks might be absent in special cases. 
To assess this, consider the following
transformation for the scalar field
\begin{equation}\label{scalartransfornation}
\tilde{\phi} = \int \frac{3[{\cal G}'_4(\phi)]^2+(1+{\cal G}_4(\phi))(1+\partial_{X}{\cal G}_{2}(\phi,X))}{(1+{\cal G}_{4}(\phi))^2}\,\mathrm{d}\phi\,.
\end{equation}
The scalar equation of motion becomes
\begin{align} \nonumber
\tilde{g}^{ab} \tilde{\nabla}_a\tilde{\nabla}_b \tilde{\phi}  = \frac{1}{(1+{\cal G}_4(\phi))^2} \Biggl\{& V'(\phi)-\partial_{\phi}{\cal G}_{2}(\phi,X)-2\,{\cal G}'_4(\phi)\frac{V(\phi)-{\cal G}_{2}(\phi,X)}{1+{\cal G}_{4}(\phi)} \\
& -{\cal G}'_{4}(\phi) \biggr[ 6{\cal G}_4''(\phi)-1+\partial_{X}{\cal G}_{2}(\phi,X)-6\frac{[{\cal G}'_4(\phi)]^2}{1+{\cal G}_{4}(\phi)}\biggr]\tilde{X} \Biggr\} \;,
\end{align}
where $\phi$ is to be understood as a function of $\tilde{\phi}$: $\phi(\tilde{\phi})$, provided the relation~\eqref{scalartransfornation} is invertible. Then, the scalar field $\tilde{\phi}$ obeys a wave equation of the original metric $\tilde{g}_{ab}$, and no pathologies would arise (unless $\tilde g^{ab}$ itself becomes singular). However, the equivalence between the new scalar field and the old one is a nontrivial question, as the transformation~\eqref{scalartransfornation} may not always be well defined. In particular, the requirement that the newly defined scalar field should verify $\tilde{\nabla}_{[\mu}\tilde\nabla_{\nu]}\tilde{\phi} = 0$ further implies that $\tilde{\nabla}_{[\mu}\left(\tilde{X}\,\partial_{\nu]}\phi\right) = 0$, thus $\tilde X \partial_a \phi$ is twist-free. Such condition could be regarded as an external constraint to ensure well posedness. In the simple example of Sec~\ref{Ex1}, we perform a similar redefinition of the scalar field which is always well defined as it does not depend on $X$. As an illustration, we show in the nonlinear example of Sec.~\ref{Ex2} how the twist evolves for several representative cases. \\

Notice that by working in the Einstein frame, we have straightforwardly recovered the conclusions 
from~\cite{Papallo:2017qvl}, i.e. {\em local well posedness} of this class of Horndeski's theories by virtue of the 
equations of motion for $g_{ab}$ and $\phi$ being symmetric hyperbolic. The question of global solutions to 
this theory is, naturally, far more involved which is not unexpected as this is already a complex question in 
General Relativity! Nevertheless, some relevant  conclusions can  be drawn. Namely,

\begin{itemize}
\item At the nonlinear level for
{\em weak data}, the equation satisfies Klainerman's {\em null condition}~\cite{2014arXiv1407.6276H} if ${\cal G}$ 
is at least
order $X$ ($\propto \tilde X$). Consequently, together with stability of Minkowski results~\cite{Christodoulou:1993uv}
or the weak null energy condition satisfaction by Einstein equations~\cite{Lindblad:2004ue}, together with
contributions of $\phi$ satisfying Strauss' conjecture \cite{2011arXiv1110.4454W}  would imply
the (subclass) of Horndeski's theories considered has a global solution in the small data case.
Beyond the weak case however, little is known;   though, as mentioned,  the propagation speed dependence on the
field and its gradient implies a high likelihood of shocks arising and/or a change in character. Would such issues arise
and be ``invisible'' to far observers? It would depend on whether they generically form {\em inside} a black hole. In such case, 
pathological issues might be shielded from problematic consequences at the classical level. A priori this seems far from
guaranteed; indeed, in the context of ref~\cite{Ripley:2019hxt},
a change in character of the equations is encountered prior to a black hole being formed. We will also illustrate such a behavior
in section \ref{Ex2}.

\item Since the speed of propagation of (perturbations of) metric tensor and scalar field can be different, black holes
are defined by the fastest outward propagation speeds. Additionally, gravitational Cherenkov
radiation would be possible and high energy cosmic rays can help to
draw constraints on this process (e.g.~\cite{Kimura:2011qn}).

\end{itemize}

Last, we can also check what we can draw from adopting the harmonic gauge in the Einstein frame and its  
implication in the Jordan frame.
For starters, it is trivial to determine that $\tilde \Gamma^a = \Omega^{-2} \, ( \Gamma^a -2 \nabla^a \ln \Omega)$.
Thus, in the Jordan frame the harmonic condition from the Einstein frame calls for adopting coordinates that
satisfy instead $\Gamma^a = 2 \nabla^a \ln \Omega$. Which implies
\begin{equation}
\Gamma^a = \frac{{\cal G}'_4}{1+{\cal G}_4} \nabla^a \phi
\end{equation}
which is precisely the condition derived in~\cite{Papallo:2017qvl} in the Jordan frame to obtain a strongly
hyperbolic system of equations and establish local well posedness.

\section{Illustration in specific cases}

\subsection{Jordan and Einstein frames equations of motion. Hyperbolicity and implications}\label{Ex1}
Within the class of Horndeski's theories, one of the simplest ones is given by,
\begin{equation}\label{jordan}
S=\frac{1}{16\pi}\int d^4x \sqrt{-g} \left[ \phi R -
\frac{\omega}{\phi} g^{\alpha\beta} \nabla_{\alpha}\phi \nabla_{\beta}\phi  \right] \; , 
\end{equation} 
where $\omega$ is a function of $\phi$ only.
A comparison with Horndeski's Lagrangian implies,
\begin{equation}
{\cal G}_2 = \frac{(2 \omega - \phi)}{\phi} X \, \, , \, \,
{\cal G}_4 = \phi - 1 \nonumber \, ;
\end{equation}
with all the other functions (including the potential) set to zero. From our previous discussion,  since $\partial_{XX} {\cal G}_2 = 0$, it is clear that in the Einstein frame characteristics
of both metric tensor and scalar field are determined by the metric. This theory has recently
been the subject of fully non-linear studies in the context of binary black neutron
star mergers~\cite{Barausse:2012da,Palenzuela:2013hsa,Shibata:2013pra}. In such scenarios
global solutions describing several orbits, merger and aftermath have been successfully achieved.
This suggests an underlying robustness of the equations of motion which can be understood at the
analytical level rather simply. To fix ideas, let us consider the vacuum case.  The field equations derived
from the (Jordan frame) action (\ref{jordan}) are
\begin{equation} \label{jordaneqmetric}
R_{\mu\nu}-\frac{1}{2} g_{\mu\nu} R= \frac{\omega}{\phi^2} \left(
\nabla_{\mu}\phi \nabla_{\nu} \phi -\frac{1}{2} g_{\mu\nu}
\nabla^{\alpha}\phi \nabla_{\alpha}\phi \right) +\frac{1}{\phi} \left(
\nabla_{\mu}\nabla_{\nu} \phi-g_{\mu\nu} \Box \phi \right)  \; , 
\end{equation}
\begin{equation} \label{jordaneqfield}
\Box \phi = -\frac{\phi R}{2 \omega} +\left(\frac{1}{2 \phi} -\frac{\omega'}{2\omega}\right)(\nabla \phi)^2 \; .  
\end{equation}
Upon replacing the Ricci scalar one re-expresses equation (\ref{jordaneqfield}) as,
\begin{equation} \label{jordaneqfieldreplacedricci}
\Box \phi = -\frac{\omega'}{3+2\omega}(\nabla \phi)^2\; .  
\end{equation}
which satisfies the null condition in the weak case. However, a non-trivial coupling 
--at the level of the principal part-- is present in equation (\ref{jordaneqmetric}). Furthermore,
notice the right hand side of this equation contains second derivatives of the scalar field --thus such terms
do belong to the principal part of the system. As well, because of such terms, the right hand side does not 
seemingly satisfy the null energy condition.
Both these observations indicate it is not a priori clear that solutions obtained from this system are well behaved. 

However, through
the conformal transformation~\cite{BD61},
\begin{equation}
\label{CT} g_{\mu\nu} \longrightarrow \tilde{g}_{\mu\nu}=\phi \,
g_{\mu\nu} \; , 
\end{equation}
 and the scalar field redefinition 
\begin{equation}
\label{SFredefinition} \phi \longrightarrow \tilde{\phi}=
\int \frac{(3+2\omega)^{1/2}}{\phi} \, d\phi \; , 
\end{equation}
 one recasts the theory in the Einstein 
frame. In this frame, the theory is defined by the
standard Einstein-Hilbert action with an extra field,

\begin{equation} \label{einstein}
S=\int d^4x
\sqrt{-\tilde{g}} \left[ \frac{\tilde{R}}{16\pi} -\frac{1}{2} \,
\tilde{g}^{\mu\nu} \tilde{\nabla}_{\mu}\tilde{\phi}
\tilde{\nabla}_{\nu}\tilde{\phi} \right] \; . 
\end{equation}
The field equations are
the usual Einstein equations with the scalar field as a source together
with a rather trivial equation for the scalar field itself,
\begin{equation}
\tilde{R}_{\mu\nu}-\frac{1}{2} \tilde{g}_{\mu\nu} \tilde{R}= 8\pi \left(
\tilde{\nabla}_{\mu}\tilde{\phi} \tilde{\nabla}_{\nu} \tilde{\phi}
-\frac{1}{2} \, \tilde{g}_{\mu\nu} \tilde{\nabla}^{\alpha} \tilde{\phi}
\tilde{\nabla}_{\alpha}\tilde{\phi} \right) \; , 
\end{equation}
\begin{equation}
\tilde{\Box}
\tilde{\phi}=0 \; .  
\end{equation}


The equation for the (conformal) metric
$\tilde g_{ab}$ is amenable to the standard analysis of well posedness in Einstein 
equations (e.g.~\cite{Sarbach:2012pr}). In particular,
adopting {\em harmonic coordinates} ($\tilde \Gamma_a = 0$) the principal part of equation (\ref{einstein}) 
becomes just ten wave equations. Further, the right hand side now obeys the null energy condition.
Thus, in the Einstein frame it follows  that at least a local in time solution will exist and standard
geometrical arguments can be exploited to assess general features of the spacetime behavior.

What does this imply in the Jordan frame? Here, since, 
$\tilde \Gamma_a = \phi^{-2} \, (\phi \Gamma_a - \nabla_a \phi)$, the discussion above suggests adopting coordinates satisfying
$\Gamma_a = \phi^{-1} \nabla_a \phi$. With this choice, the equations of motion in the
Jordan frame can be re-expressed in the following way.
Beginning with
\begin{equation}
R_{ab} = \frac{\omega}{\phi^2} \nabla_a \phi \nabla_b \phi + \frac{1}{2 \phi} g_{ab} \Box \phi + \frac{1}{\phi} \nabla_a \nabla_b \phi\,,
\end{equation}
we then define $\hat R_{ab} + \nabla_{(a} \Gamma_{b)} \equiv R_{ab}$ (i. e. taking out the covariant derivative of the
trace of the Christoffels). Now, replacing in such a term the condition on the coordinates, we obtain
\begin{equation}
\hat R_{ab} = \frac{\omega+1}{\phi^2} \nabla_a \phi \nabla_b \phi + \frac{1}{2} g_{ab} \phi^{-1} \Box \phi \,.
\end{equation}
A priori we still have second order derivatives in the right hand side of the above equation, but --on shell-- we can use
the equation for the field $\phi$ still. Recall,
\begin{equation}
\Box \phi = -\frac{\omega'}{(3+2\omega)}(\nabla \phi)^2\,.
\end{equation}
Thus, the metric equation results in
\begin{equation}
\hat R_{ab} = \frac{(1+\omega)}{\phi^2} \nabla_a \phi \nabla_b \phi -\frac{\omega'}{2\phi(3+2\omega)}g_{ab} (\nabla \phi)^2\,. \label{newjordan}
\end{equation}
And it is evident the right hand side can satisfy the null energy condition for $w \ge 1$.

\subsection{Exploring the non-linear behavior. Case with $\partial_{XX} {\cal G}_2 \neq 0$}\label{Ex2}
We now turn our attention to Horndeski's theories with a nonlinear kinetic term ${\cal G}_2(\phi, X) = - g X^2$, with all other functions, as well as the potential, set to zero for simplicity. This choice, similar to those adopted
in \citep{Akhoury:2011hr}, can be thought of as the first nonlinear term in a Taylor expansion of the kinetic term in a $k$-essence theory,

\begin{equation}
S = \int d^4 x \sqrt{-g} \left[ R + X - g X^2\right] \label{G2action}\, .
\end{equation}

Our goal is to study the nonlinear behavior of the theory and explore the possible phenomenology that
can arise. While we are restricting to a rather special case, as we shall see, a number of possible
pitfalls can appear which are likely to manifest in more general cases. To simplify the treatment and 
presentation, we concentrate on spherically symmetric scenarios and present several cases
defined by different initial conditions as well as the value of the coupling $g$.
For simplicity we adopt Schwarzschild coordinates where the metric can be written as,
\begin{equation}
ds^2 = - \alpha^2 dt^2 + a^2 dr^2 + r^2 d\Omega^2\, \label{G2metric}.
\end{equation}
Thus the only dynamical metric functions are the lapse function $\alpha(t, r)$
and $a(t,r)$. Recall that these coordinates become singular when a horizon forms. Such scenario
takes place when  $l^\mu \nabla_\mu r = 0$, where $l^\mu$ is a null vector \citep{Akhoury:2011hr}. In the gauge (\ref{G2metric}), this is simply $\alpha = 0$. Consequently, with our current implementation we
can explore up to black hole formation. Despite this limitation, as we shall see below, one can
identify several problematic scenarios arising either outside the black hole or even prior to its
formation. Thus, severe restrictions to well posedness arise which are not cloaked by a horizon for
asymptotic observers.

To simplify the discussion and the numerical implementation, we introduce standard first order variables
as used in \citep{Choptuik:1992jv},

\begin{equation}
\Phi \equiv \phi'\, , \qquad \Pi \equiv \frac{a}{\alpha} \dot \phi \label{G2firstorderdef}\, ,
\end{equation}
using the notation $\dot f = \partial_t f$ and $f' = \partial_r f$. In the special case of ${\cal G}_2(\phi, X) = {\cal G}_2(X)$, as in (\ref{G2action}), equations (\ref{einsteingabpregauge}) and (\ref{einsteinphipregauge}), respectively, take the form

\begin{equation}
 R_{\mu \nu} - \frac12 g_{\mu \nu} R = \left[\frac{X + {\cal G}_2(X)}{2}\right] g_{\mu \nu} +   \left[\frac{1 + \partial_X {\cal G}_2(X)}{2}\right] \nabla_\mu \phi \nabla_\nu \phi \label{G2gabeom}\, ,
\end{equation}

\begin{equation}
\left[g^{\mu\nu} - \frac{\partial^2_{XX} {\cal G}_2(X)}{1 + \partial_X {\cal G}_2(X)} \nabla^\mu \phi \nabla^\nu \phi \right] \nabla_\mu \nabla_\nu \phi = 0 \label{G2phieom}
\end{equation}
where the effective inverse metric $\gamma^{\mu \nu}$, as in equation (\ref{effectiveinversemetric}), is given by

\begin{equation}
\gamma^{\mu \nu} =  g^{\mu\nu} - \frac{\partial^2_{XX} {\cal G}_2(X)}{1 + \partial_X {\cal G}_2(X)} \nabla^\mu \phi \nabla^\nu \phi  \label{G2gamma}\, .
\end{equation}

Now, in order to monitor the character of the equation of motion for the scalar field (\ref{G2phieom}), the eigenvalues of the effective inverse metric must be computed. In particular we extract at any given time the two eigenvalues, here labeled as $\lambda_\pm$ for every spatial point. Since we are mainly interested in one of the eigenvalues going to zero, the relevant quantities will be $\min(\lambda_+)$ and $\max(\lambda_-)$, where $\min(\cdot)$ and $\max(\cdot)$ refer to the minimum and maximum in the spatial (radial) direction, at any given time. It is important to keep in mind that although $\lambda_+ > 0$ and $\lambda_- < 0$ for $\phi = 0$, this is not necessarily the case for arbitrary configurations. In fact, the equations will change character when these conditions cease to be satisfied. The two eigenvalues can be expressed as

\begin{equation}
\lambda_\pm = \frac{\gamma^{tt} + \gamma^{rr}}{2} \pm \sqrt{\left(\frac{\gamma^{tt} + \gamma^{rr}}{2}\right)^2 - \gamma^{tt}\gamma^{rr} + (\gamma^{tr})^2} = \frac{\gamma^{tt} + \gamma^{rr}}{2} \pm \sqrt{\left(\frac{\gamma^{tt} + \gamma^{rr}}{2}\right)^2 - \det (\gamma^{\mu\nu})}\label{lambdaeigs}\; .
\end{equation}

It is evident that the system will become parabolic when $\det (\gamma^{\mu\nu}) = 0$, as expected. Additionally, it is important to keep track of the characteristic speeds, or propagation velocities, of the scalar field. This can be done by extracting the eigenvalues, here labeled as $V_\pm$, of the principal part of the (first order) equations of motion for $\Phi$ and $\Pi$. These eigenvalues determine the shape of the light cones for the scalar field, and can be used to identify features such as sound horizons (horizons for the scalar field \citep{Akhoury:2011hr}). With our conventions, asymptotically $V_+ \rightarrow 1$ while $V_- \rightarrow -1$ describing, respectively, the incoming and outgoing modes of the field.
A sound horizon --with respect to asymptotic observers-- will appear\footnote{Naturally the opposite
condition still defines a local sound horizon, cloaking some local region from being reached by
scalar field perturbations.} when $V_- = 0, V_+ \ge 0$. Again, as in the case of the effective metric, we are interested in $\min(V_+)$ and $\max(V_-)$.

\begin{equation}
V_\pm = - \frac{\gamma^{tr}}{\gamma^{tt}} \pm \sqrt{\left(\frac{\gamma^{tr}}{\gamma^{tt}}\right)^2 - \frac{\gamma^{rr}}{\gamma^{tt}}}  = - \frac{\gamma^{tr} }{\gamma^{tt}} \pm \sqrt{-\frac{\det(\gamma^{\mu\nu})}{(\gamma^{tt})^2}}\label{Veigs}\, .
\end{equation}

As mentioned, when $\det (\gamma^{\mu\nu}) = 0$ the equation changes character. However, the rate
at which $(\gamma^{tt})^2 \rightarrow 0$ distinguishes two important cases with respect
of the {\em type} of change. Recall that mixed character equations can often be classified in comparison 
to two standard equations (\cite{Ripley:2019irj}). These are the Tricomi equation
\begin{equation}
\partial_y^2 u(x, y) + y \partial_x^2 u(x, y) = 0  \label{Tricomi}\, ,
\end{equation}
where the characteristic speeds, $\pm{{y}^{1/2}}$, go to zero at the character transition line $y=0$, and the Keldysh equation
\begin{equation}
\partial_y^2 u(x, y) + \frac{1}{y} \partial_x^2 u(x, y) = 0  \label{Keldysh}\, ,
\end{equation}
where the speed $\pm{{y}^{-1/2}}$ diverges at the transition line. 

Notice that  the discriminant between the two characteristic speeds (\ref{Veigs}) turns out to be proportional 
to $-\det(\gamma^{\mu\nu})$. Therefore,
as long as $(\gamma^{tt})^2 \rightarrow 0$ slower than $\det(\gamma^{\mu\nu}) \rightarrow 0$, the characteristic speeds $V_+,V_-$ will coincide and the scalar field light cone becomes degenerate. Thus, there must exist some instant of time, before the system becomes --at least locally-- parabolic, when either $V_+$ or $V_-$ is zero (the latter case implying a sound horizon) indicative
of a Tricomi-type transition. On the other hand, if
$(\gamma^{tt})^2 \rightarrow 0$ faster than $\det(\gamma^{\mu\nu}) \rightarrow 0$ the characteristic speeds diverge indicating a transition of Keldysh type. This case is more delicate to tract numerically as
the diverging speeds imply the time-step should be adjusted to decrease inversely with the maximum speed with
an explicit integration algorithm. (Note: an implicit update could be implemented to bypass this issue,
but at the expense of missing physics taking place at smaller scales than the time-step adopted).

Interestingly, in \cite{Ripley:2019irj}, only a Tricomi-type behavior is observed. Anticipating our results,  we observe both cases depending on the value of the coupling $g$: Tricomi-like for $g < 0$ and Keldysh-type transitions $g > 0$. The well posedness of Tricomi equation has been explored in \cite{doi:10.1002/cpa.3160230404,Otway} and, as discussed in~\cite{Ripley:2019irj}
the initial/boundary conditions to ensure well posedness would be rather unnatural from a time-development point of view.

\subsubsection*{Implementation details}

In the first order variables (\ref{G2firstorderdef}) we can extract from the $rr$ and $tt$ components of equation (\ref{G2gabeom}), respectively, the first order constraint equations 

\begin{equation}
\alpha' = \frac{\alpha}{8r} \left[4(a^2 - 1) + r^2 (\Phi^2 + \Pi^2) \right] -  g\frac{r \alpha}{16 a^2} \left[(\Phi^2 + \Pi^2)^2 - 4\Phi^4\right]\label{G2alpha}\, ,
\end{equation}

\begin{equation}
a' = \frac{a}{8r} \left[4(1 - a^2) + r^2 (\Phi^2 + \Pi^2) \right] + g\frac{r}{16 a}  \left[(\Phi^2 + \Pi^2)^2 - 4\Pi^4\right] \label{G2a}\, .
\end{equation}

Equation (\ref{G2phieom}), in terms of the first order variables, is given by

\begin{align}
\dot \Pi =
\frac{1}{r^2}\left(r^2 \frac{\alpha}{a} \Phi\right)'
+ \frac{2 g}{a^2 + g \left(\Phi^2 - 3 \Pi^2\right)} \frac{\alpha}{a}  
  \Bigg[   (\Phi^2 + \Pi^2) \Phi ' - 2 \Phi \Pi  \Pi'  
 \nonumber \\ +  \left( \frac{r}{4}\Pi ^2 - \frac{a'}{a}\right) \left(\Phi^2-\Pi ^2\right)\Phi
+ \frac{g r}{4 a^2} \left(\Phi^2- \Pi^2\right)^2 \Phi  \Pi ^2
+ \frac{2}{r} \Phi \Pi ^2 \Bigg] \label{G2Pi}\, ,
\end{align}
together with the condition that $\partial_t \partial_r \phi = \partial_r \partial_t \phi$, namely
\begin{equation}
\dot \Phi =  \left(\frac{\alpha}{a}\Pi\right)' \label{G2Phi}\, .
\end{equation}

The effective inverse metric from equation (\ref{G2gamma}) reads 

\begin{equation}
\gamma^{tt} = -\frac{1}{\alpha ^2} \left( 1-g \frac{2 \Pi ^2}{a^2+g \left(\Phi ^2 -\Pi ^2 \right)}\right)\, , \qquad \gamma^{rr} = \frac{1}{a^2} \left(1 + g\frac{2 \Phi ^2}{a^2+g \left(\Phi ^2-\Pi ^2\right)} \right)\, ,
\end{equation}
\begin{equation}
\gamma^{tr} = -g\frac{2 \Pi  \Phi }{a \alpha  \left(a^2+g \left(\Phi ^2-\Pi ^2\right)\right)}\, ,
\end{equation}

and the matrix defining the principal part of equations (\ref{G2Pi}) and (\ref{G2Phi}) is,

\begin{equation}
M = \left(
\begin{array}{cc}
 0 & \frac{\alpha}{a} \\
 -\frac{a \gamma^{rr}}{\alpha \gamma^{tt}} &  - 2 \frac{\gamma^{tr}}{\gamma^{tt}} \\
\end{array}
\right) =
\frac{\alpha}{a}\left(
\begin{array}{cc}
 0 & 1 \\
 1 + 2g\frac{\Phi ^2 + \Pi ^2}{a^2 + g \left(\Phi ^2-3 \Pi ^2\right)} & -4 g \frac{ \Pi  \Phi }{a^2 + g \left(\Phi ^2-3 \Pi ^2\right)} \\
\end{array}
\right)\; .
\end{equation}

The equations of motion are solved in a constrained evolution scheme. Both $\alpha(t, r)$ and $a(t, r)$ are obtained through a spatial integration while the scalar field is integrated in time through a Runge Kutta 4th order time integrator. At each time step (intermediate or full), given a spatial profile for the fields $\Phi$ and $\Pi$, the constraint equations (\ref{G2alpha}) and (\ref{G2a}) are integrated in space using also a Runge-Kutta 4th order method (RK4). First, $a$ is integrated radially outwards from $r = 0$ to $r = r_\text{max}$ with the initial condition $a(r = 0) = 1$. This condition ensures regularity at ($\alpha' = a' = 0$) at the origin. Then, $\alpha$ is integrated radially inwards with the condition $\alpha(r_\text{max}) = 1 / a(r_\text{max})$. Notice that, as these integrations are carried out, the fields $\Phi$ and $\Pi$ which are needed at `virtual radial points' in between grid points are obtained through fourth order (second order near the spatial boundaries) spatial interpolations at any given time step. 

Evolution of $\Phi$ and $\Pi$ forward in time is carried out via the method of lines with a RK4 integration using equations (\ref{G2Pi}) and (\ref{G2Phi}). Spatial derivatives are computed with second order (first order near the boundaries) finite-difference operators satisfying summation by parts. Regularity at the origin is addressed by using l'H\^opital's rule at $r=0$ to
regularize the equation, and we defined totally outgoing boundary conditions at the outer radial boundary. A small amount of fourth order (second order near the boundaries) artificial dissipation is added for convenience as described. For further details see \cite{Calabrese:2003yd,Calabrese:2003vx,Guzman:2007ua}. 

The numerical results displayed in this paper are performed in a spatial domain ranging from $r = 0$ to $r = r_\text{max} = 100$, and a spatial resolution of $\Delta r = 1/80$. (though convergence and consistency of the solutions obtained is checked with resolutions
of $\Delta r = 1/20$ and $\Delta r = 1/40$). The Courant number is initially taken to be $C = 1/10$, and therefore $\Delta t = C \Delta r = 1/800$. Numerical output is produced every 40 time steps. For cases 
displaying very fast changes, or a high speed of propagation of the scalar field, we switch to a Courant parameter of $C = 1/100$ ($\Delta t = 1/8000$) in the last part of the simulation, and we produce output of
the solution every 4 time steps. For reference, the times when these refinements are initiated are listed in the Appendix.

\subsubsection*{Initial conditions and coupling parameters}
As mentioned, our goal is to explore the possible phenomenology that can arise
in this theory. We have performed extensive studies to try and isolate different scenarios and,
for concreteness in our presentation, we present three representative cases for positive and negative coupling
values.  In particular, we adopt initial data for the (first order variables of the) scalar field
given by:

\begin{equation}
\Phi(t = 0, r) = A \exp \left( - \frac{(r - r_0)^2}{\sigma^2}\right) \cos \left( \frac{\pi}{10} r \right)\, , \qquad \Pi(t = 0, r) = 0\, .
\end{equation}

with $r_0 = 55$. The three cases, labeled {\bf A}, {\bf B} and {\bf C}, are defined by the following
parameters:

\begin{itemize}
\item Case {\bf A}: $A = 0.02$, $\sigma = 15.0$
\item Case {\bf B}: $A = 0.14$, $\sigma = 1.5$
\item Case {\bf C}: $A = 0.045$, $\sigma = 15.0$
\end{itemize}

For each of these parameter sets, we have obtained solutions for $g = +1$ (labeled A+, B+ and C+) and for $g = -1$ (A-, B- and C-).  Naturally, the scale over which a non-trivial physical behavior occurs depends on: (i) the initial location and amplitude of the pulse --as it travels towards the origin in spherical symmetry, his associated energy density naturally grows-- and (ii) the strength of the coupling parameter $g$.

\subsection*{Negative coupling constant: $g = -1$}

\begin{figure}[h!]
\includegraphics[width=150pt]{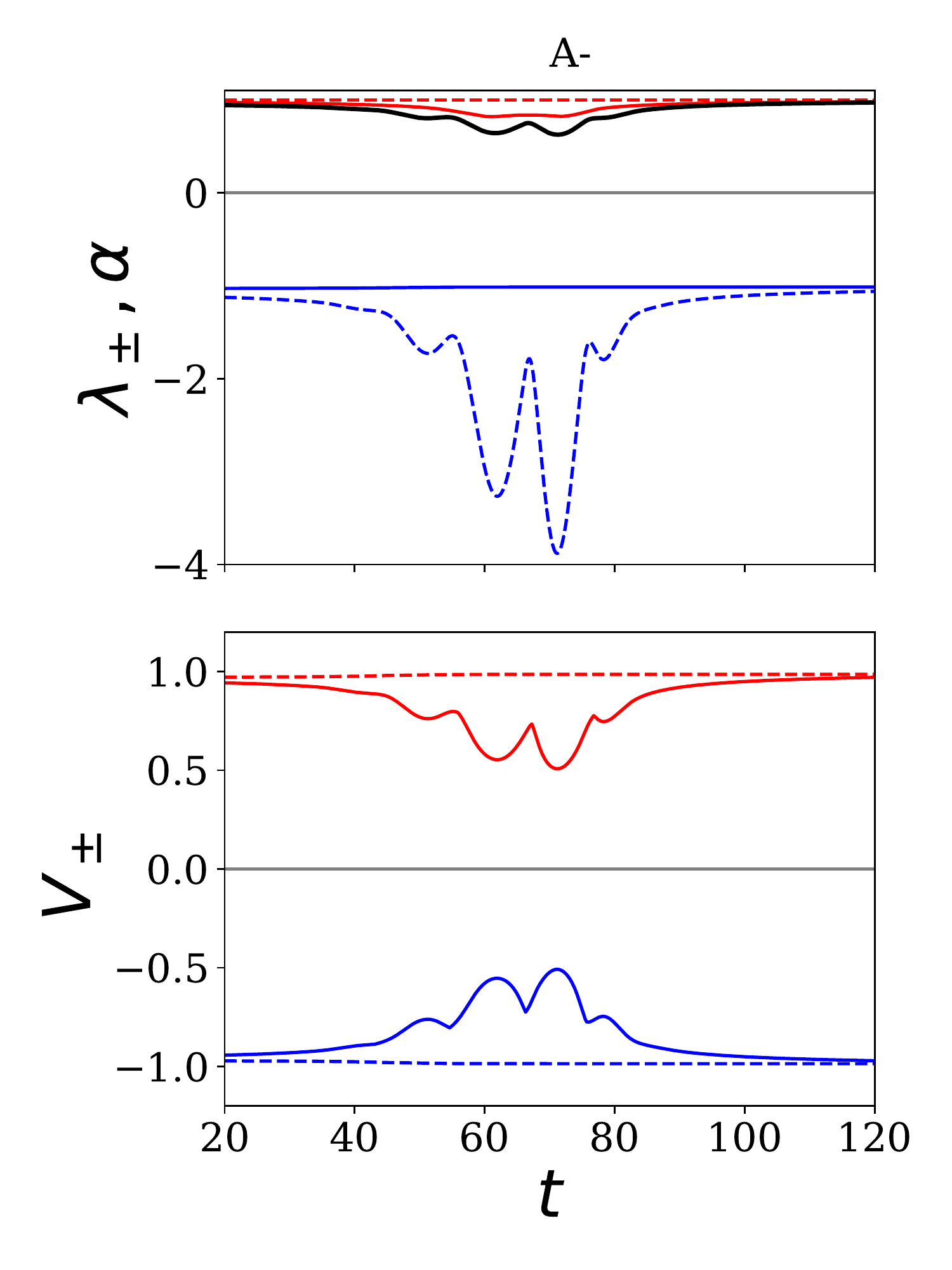}
\includegraphics[width=150pt]{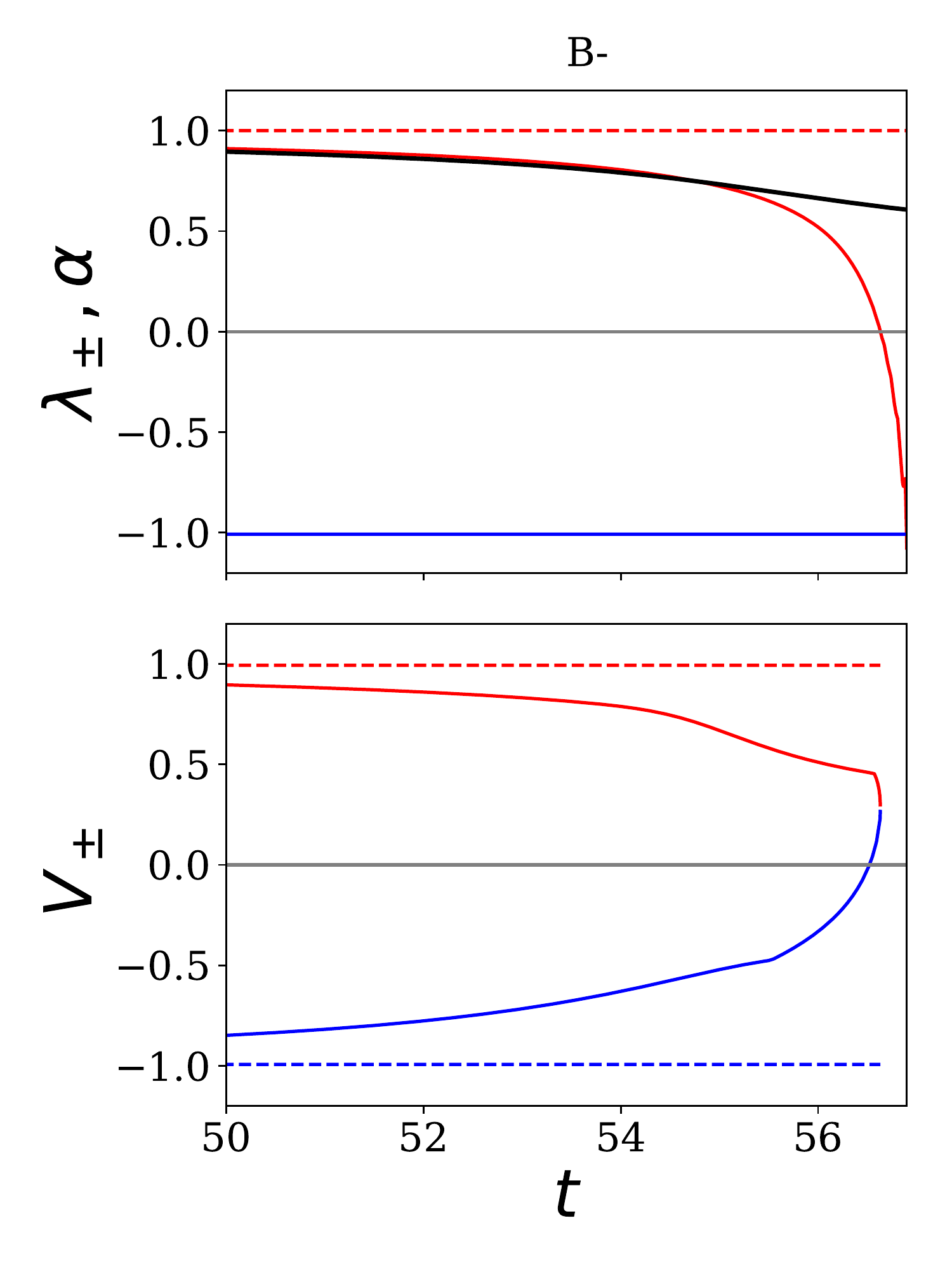}
\includegraphics[width=150pt]{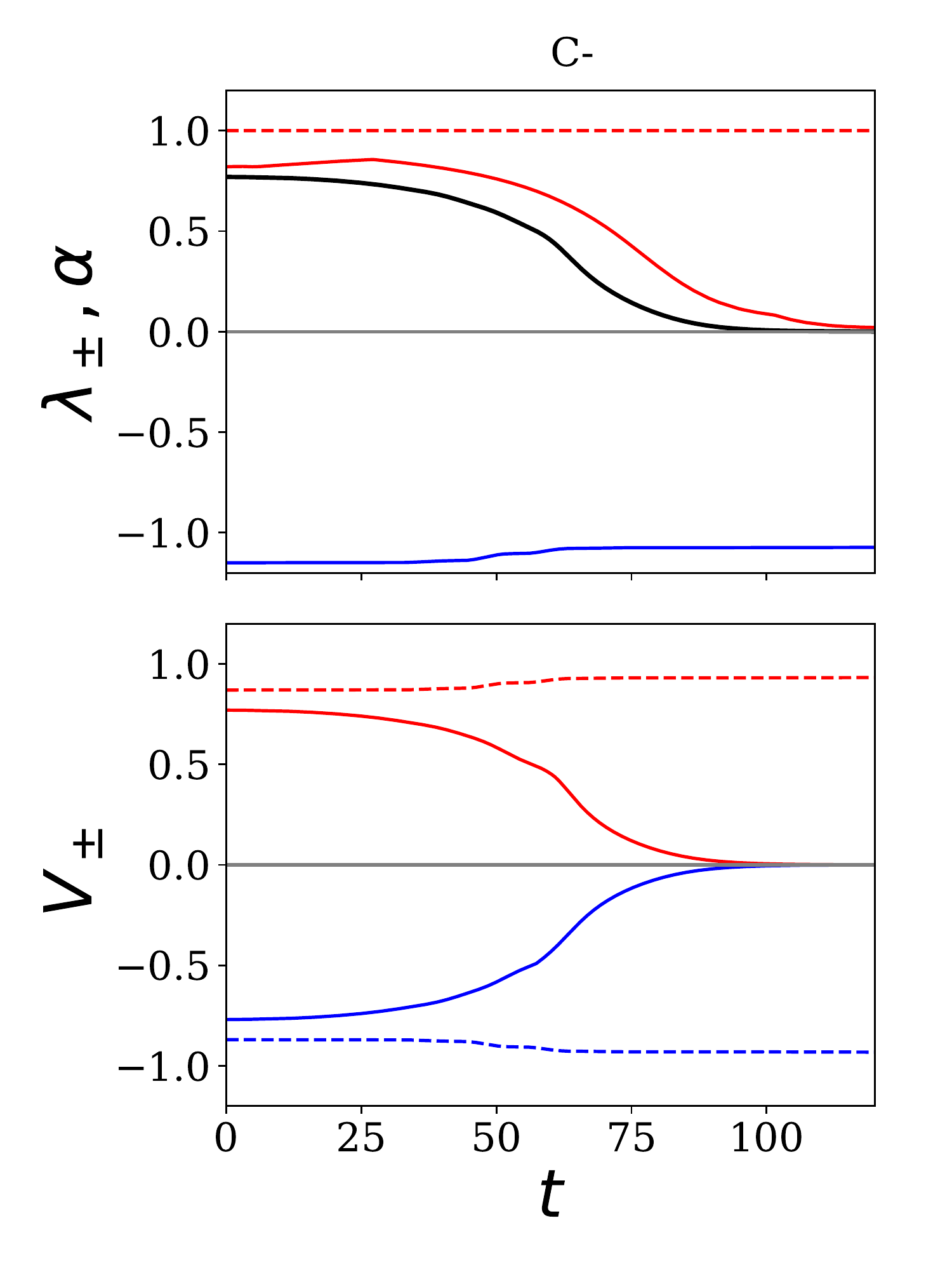}
\caption{Eigenvalues for $g = -1$, in cases A- (left), B- (center) and C- (right).  The upper three plots show the (max/min of)
 eigenvalues $\lambda_\pm$ of the effective inverse metric $\gamma^{\mu\nu}$ and the minimum of $\alpha$. The lower three plots show the eigenvalues $V_\pm$ of the principal part of the scalar field equations, corresponding to the characteristic speeds of propagation of the scalar. In each plot, the upper red curves correspond to the spatial maximum (red dashed) and minimum (red solid) values of the $\lambda_+$ and $V_+$, while the lower blue curves depict the spatial maximum (blue solid) and minimum (blue dashed) of $\lambda_-$ and $V_-$. The thick black solid line is the lapse function $\alpha$, used to identify the formation of a black hole. A gray line at 0 is added as a guide to the eye.\label{fig:EigenvaluesMinus}}
\end{figure}

Setting $g = -1$, we observe three different outcomes depending on the initial conditions of the wave pulse as illustrated in FIG. \ref{fig:EigenvaluesMinus}. If the data is weak enough, case A-, the ingoing pulse reaches the origin, bounces off it and disperses as it propagates to infinity.  For configuration B-, the eigenvalue $\lambda_+$ of the effective inverse metric crosses zero at $t \approx 56.63, r \approx 1.75$ while the lapse remains bounded from below by $\alpha \approx 0.62$. This indicates the system has become parabolic before a light horizon forms. Further, as predicted by equation (\ref{Veigs}), the characteristic speeds of the scalar field merge together as $\lambda_+ \to 0$ and acquire an imaginary part after that. Before the transition point, the eigenvalue $V_-$ crosses zero at $t \approx 56.52, r \approx 1.90$, and therefore a sound horizon is indeed produced. However, since the lapse function $\alpha$ is positive everywhere, there is no light horizon and perturbations of the metric tensor can still propagate through the sound horizon, thus the transition point is not disconnected from outside observers. This is not the only possible outcome for strong enough initial data, as in configuration C- a light horizon does form, together with a sound horizon at $r \approx 6.5$, without
any change in character of the scalar field equation. In FIG. \ref{fig:EigenvaluesMinus} C-, the final state at and outside this region is described by a black hole with an outwards propagating field. As mentioned, we can not comment on what takes place inside the horizon. Interestingly, case B- displays characteristic speeds going to zero before going imaginary where the equation changes character to parabolic. This, as discussed in \cite{Ripley:2019irj}, is an indication that the equation is of Tricomi type.

\subsection*{Positive coupling constant: $g = +1$}

\begin{figure}[h!]
\includegraphics[width=150pt]{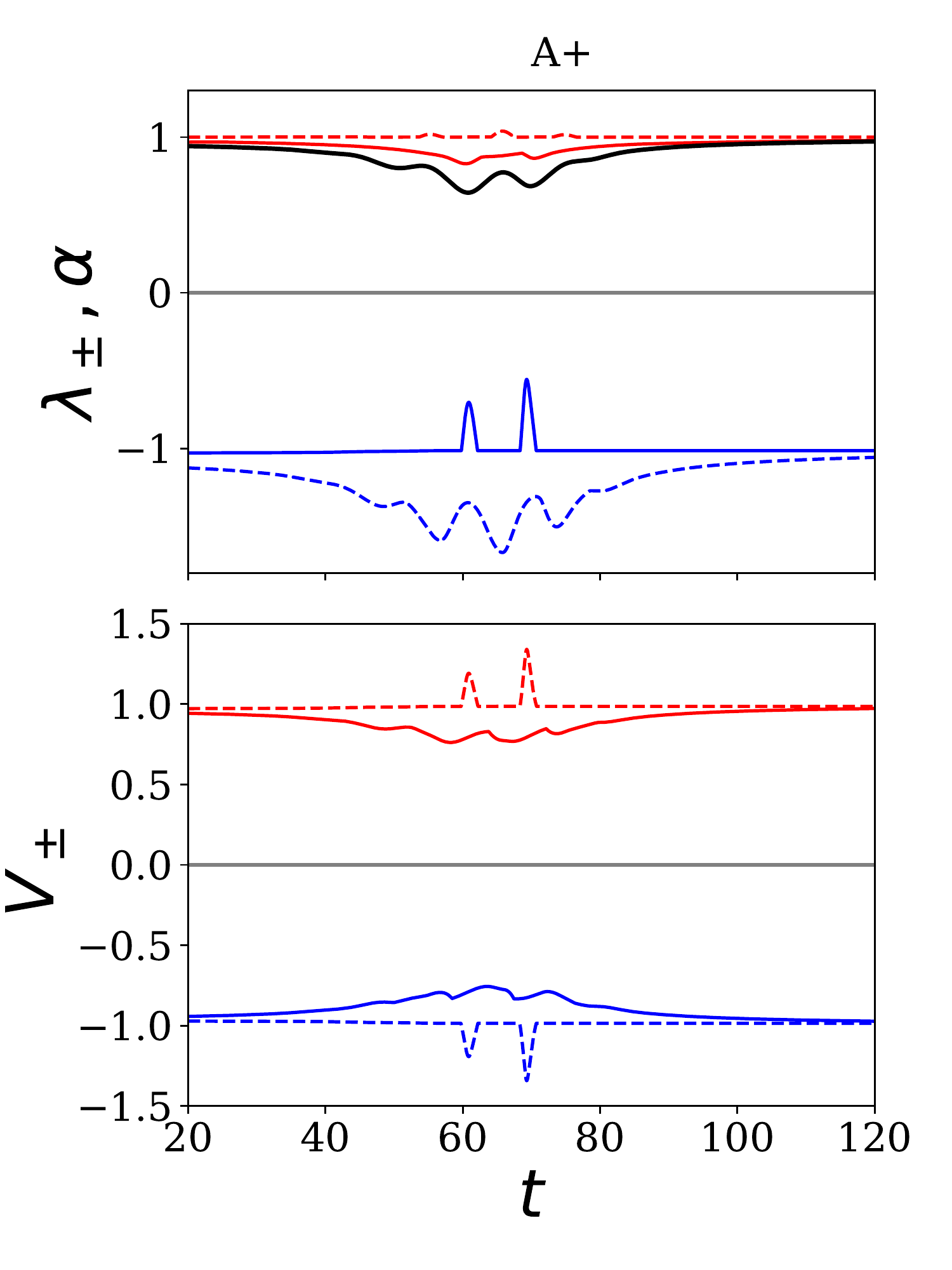}
\includegraphics[width=150pt]{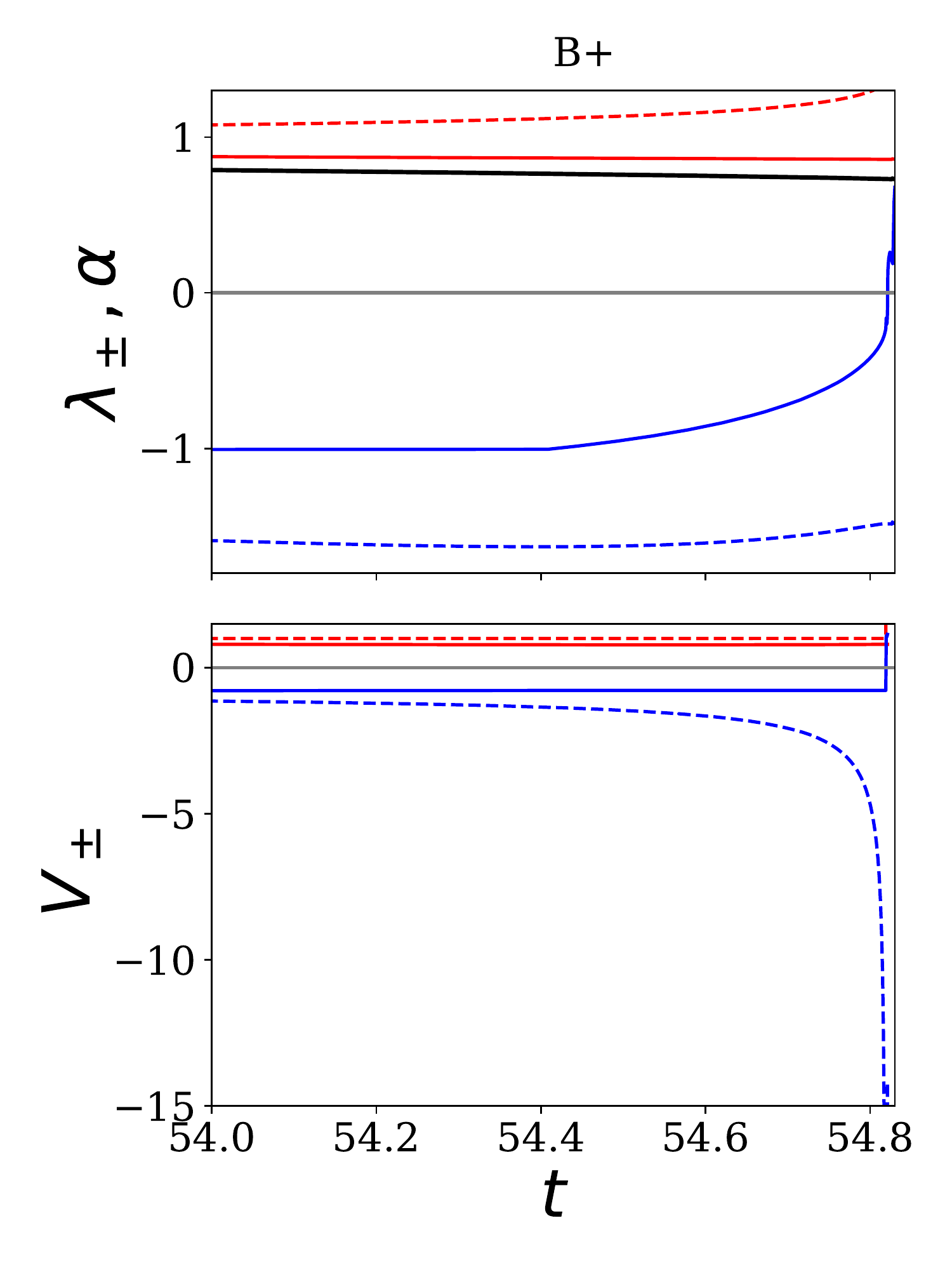}
\includegraphics[width=150pt]{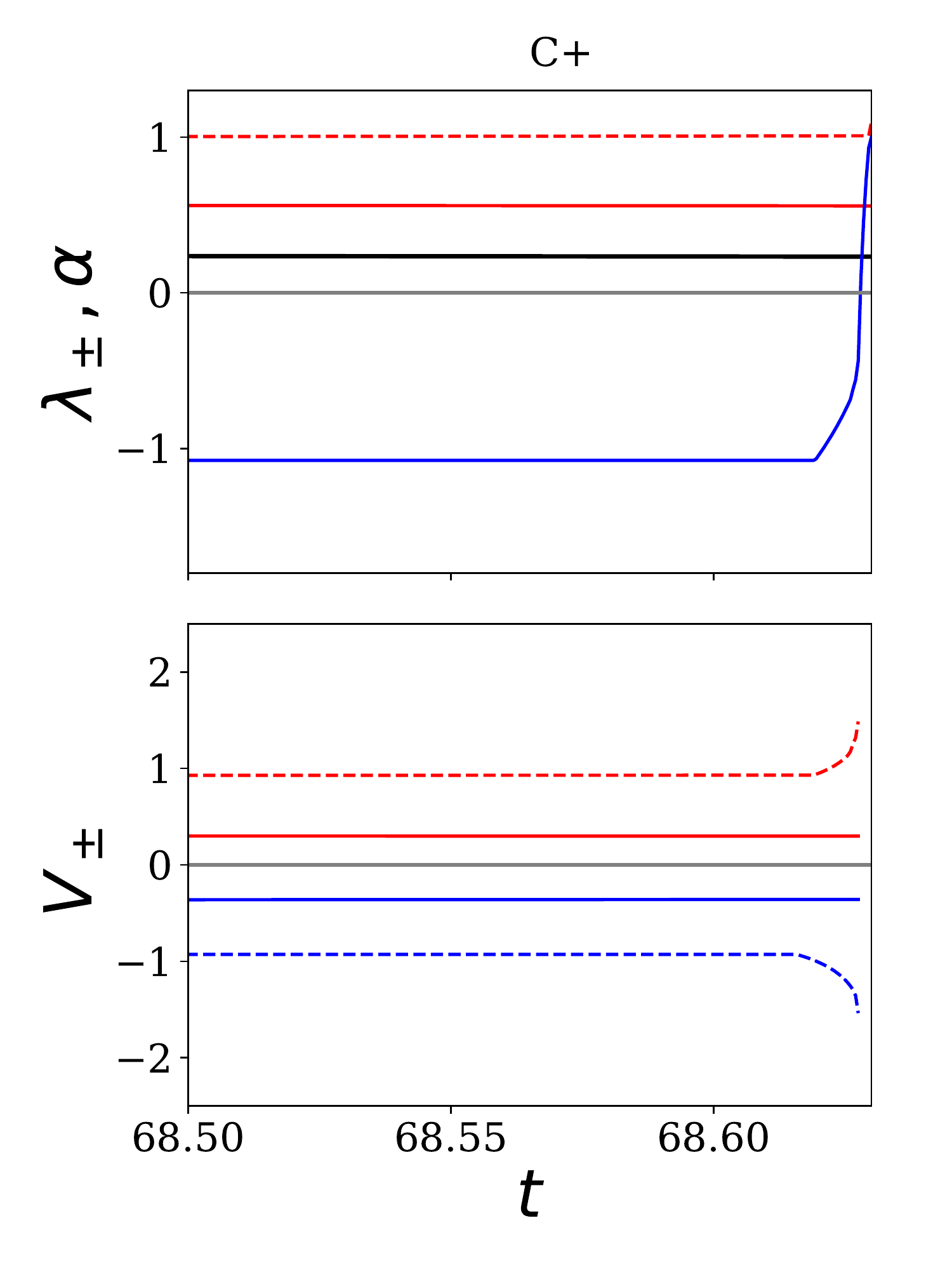}
\caption{Eigenvalues for $g = +1$, in cases A+ (left), B+ (center) and C+ (right).  The upper three plots show the (max/min of)
 eigenvalues $\lambda_\pm$ of the effective inverse metric $\gamma^{\mu\nu}$ and the minimum of $\alpha$. The lower three plots show the eigenvalues $V_\pm$ of the principal part of the scalar field equations, corresponding to the characteristic speeds of propagation of the scalar. In each plot, the upper red curves correspond to the spatial maximum (red dashed) and minimum (red solid) values of the $\lambda_+$ and $V_+$, while the lower blue curves depict the spatial maximum (blue solid) and minimum (blue dashed) of $\lambda_-$ and $V_-$. The thick black solid line is the lapse function $\alpha$, used to identify the formation of a black hole. A gray line at 0 is added as a guide to the eye.\label{fig:EigenvaluesPlus}}
\end{figure}

For $g = +1$,  delicate features in the solution for the same initial conditions developed in a more marked way and, arguably, more violently. The obtained behavior is illustrated in FIG. \ref{fig:EigenvaluesPlus}. Naturally, there is not much qualitative difference in A+ configuration. This is to be expected since for weak enough data, the impact of the scalar field is considerably suppressed. In cases B+ and C+, however, $\lambda_-$ crosses zero and the system becomes parabolic in a rather sharp, abrupt way. The transition occurs at $t \approx 54.82, r \approx 1.70$ for case B+, and at $t \approx 68.63, r \approx 0$ for case C+.

In contrast to the previous case, cases B+, C+ display fastly growing characteristic speeds right before becoming imaginary where the equation changes character to parabolic. This, as discussed in~\cite{Ripley:2019irj}, is an indication that the equation is of Keldysh type. Moreover, this implies these regimes have a natural causal horizon significantly larger than that of light (e.g.~\cite{Bonvin:2006vc}). Nevertheless, the change of character in the equation signals well-behaved solutions can only be obtained within a finite range of time. Furthemore, this change of character --for both values
of coupling-- takes place {\em prior to a shock being formed}.

\begin{figure}[h!]
\includegraphics[width=150pt]{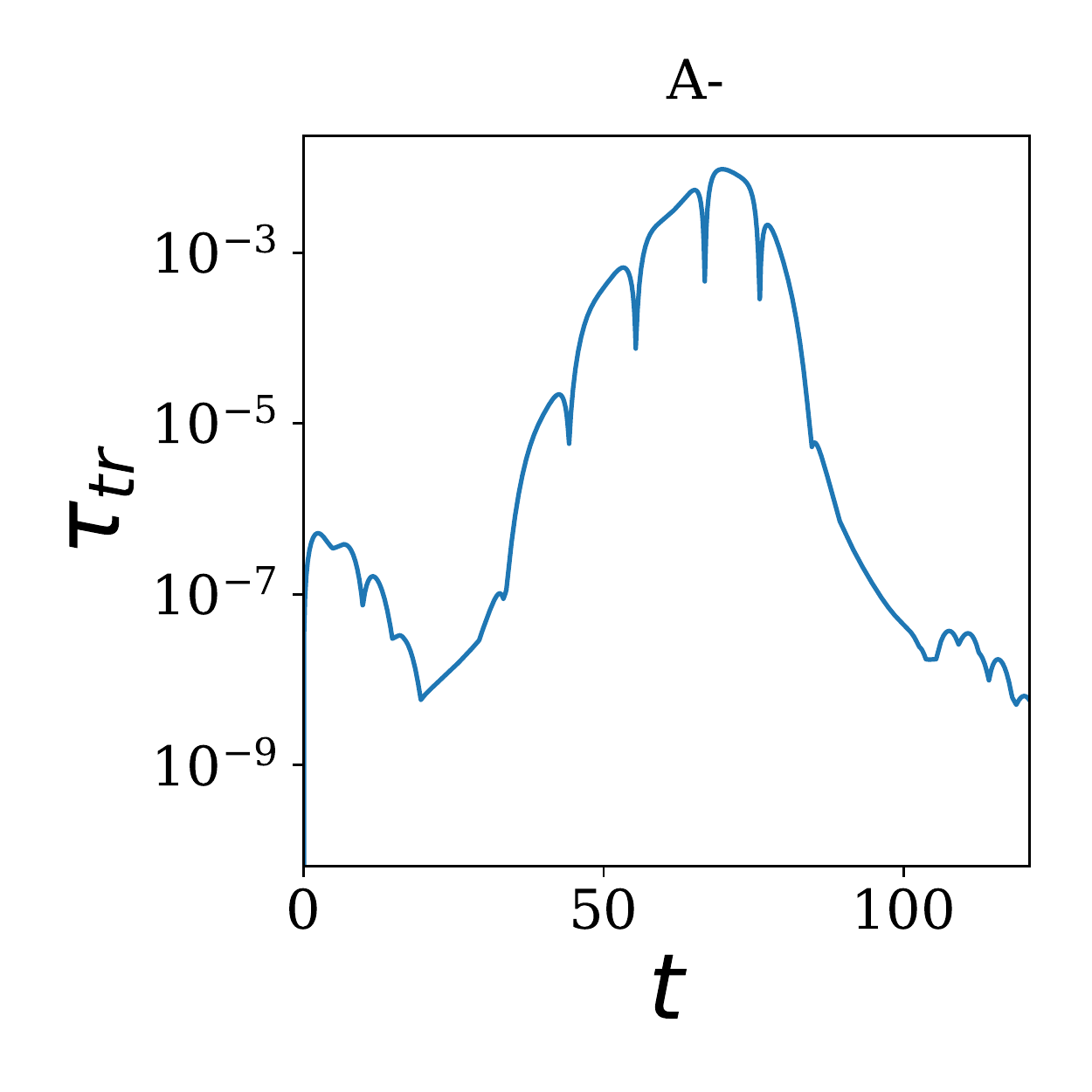}
\includegraphics[width=150pt]{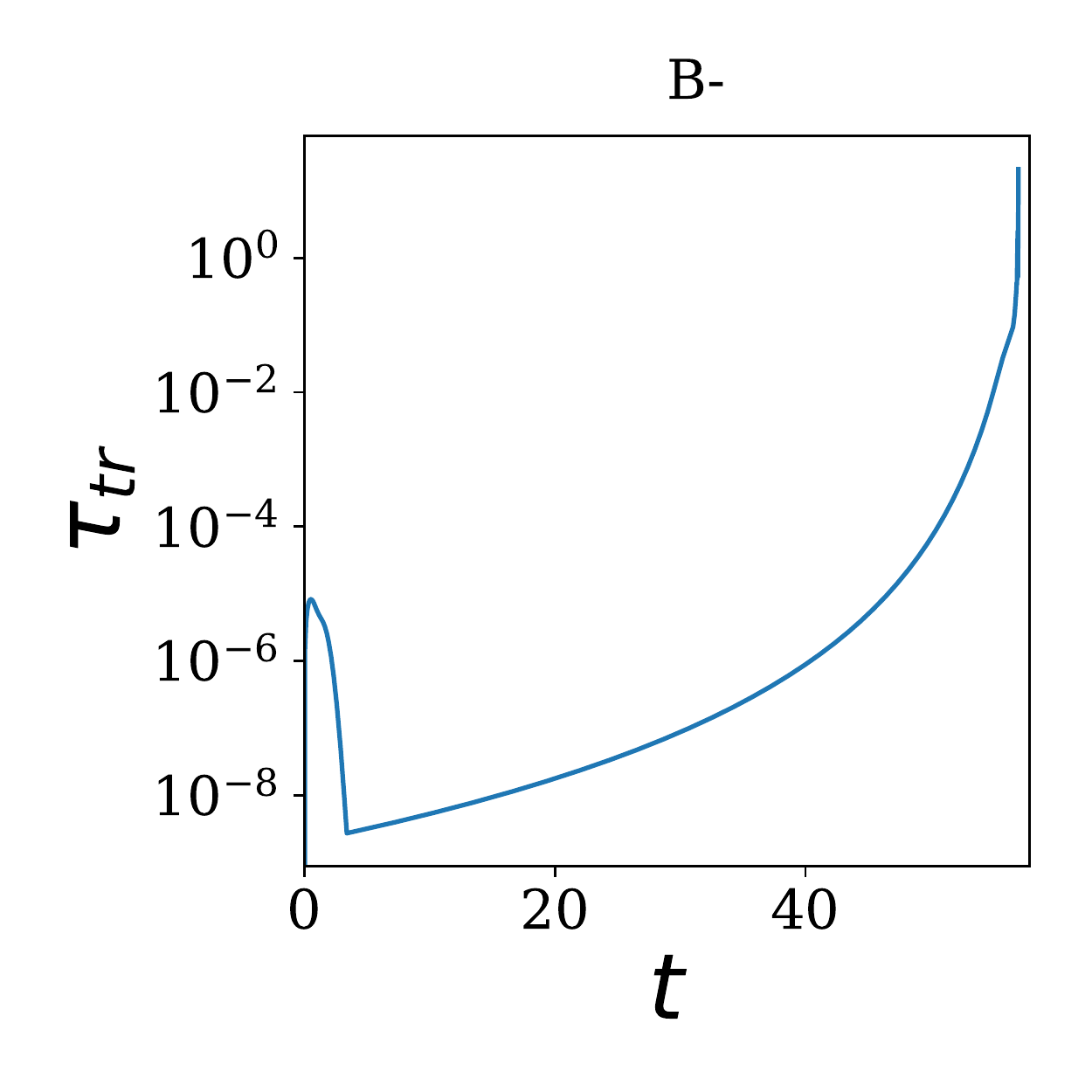}
\includegraphics[width=150pt]{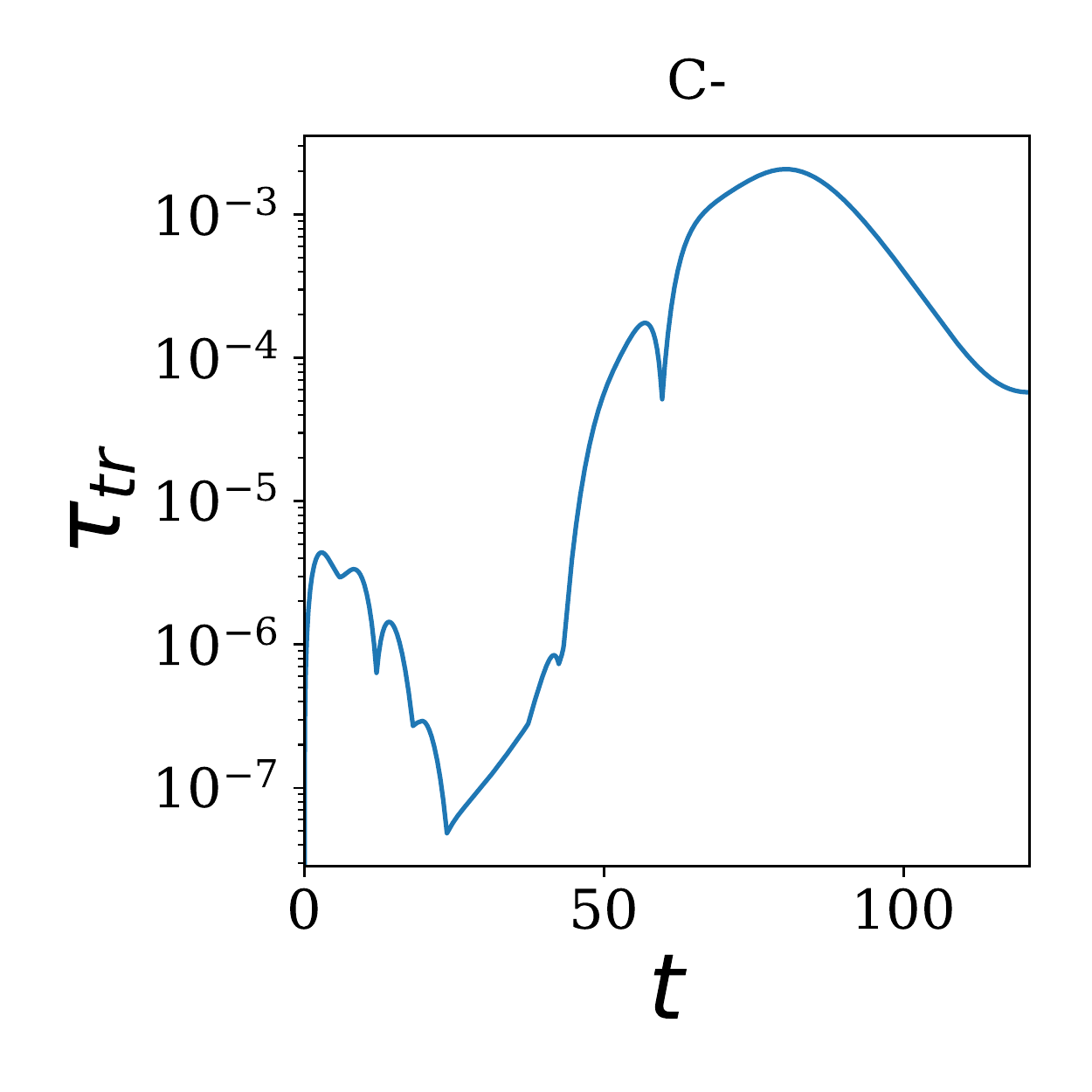}
\caption{$\max|\tau_{tr}|$ for cases A- (left), B- (center) and C- (right).\label{fig:TwistMinus}}
\end{figure}

\begin{figure}[h!]
\includegraphics[width=150pt]{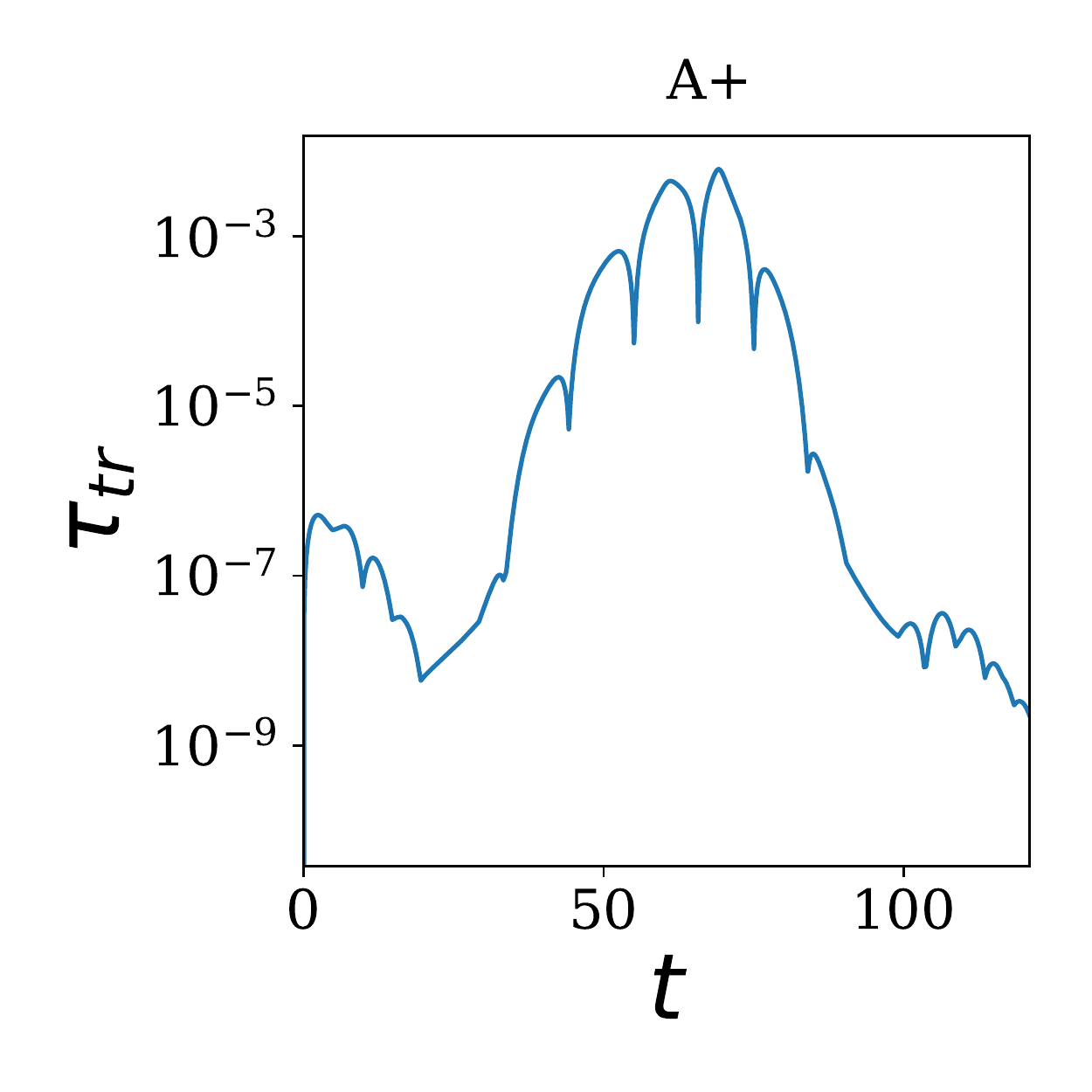}
\includegraphics[width=150pt]{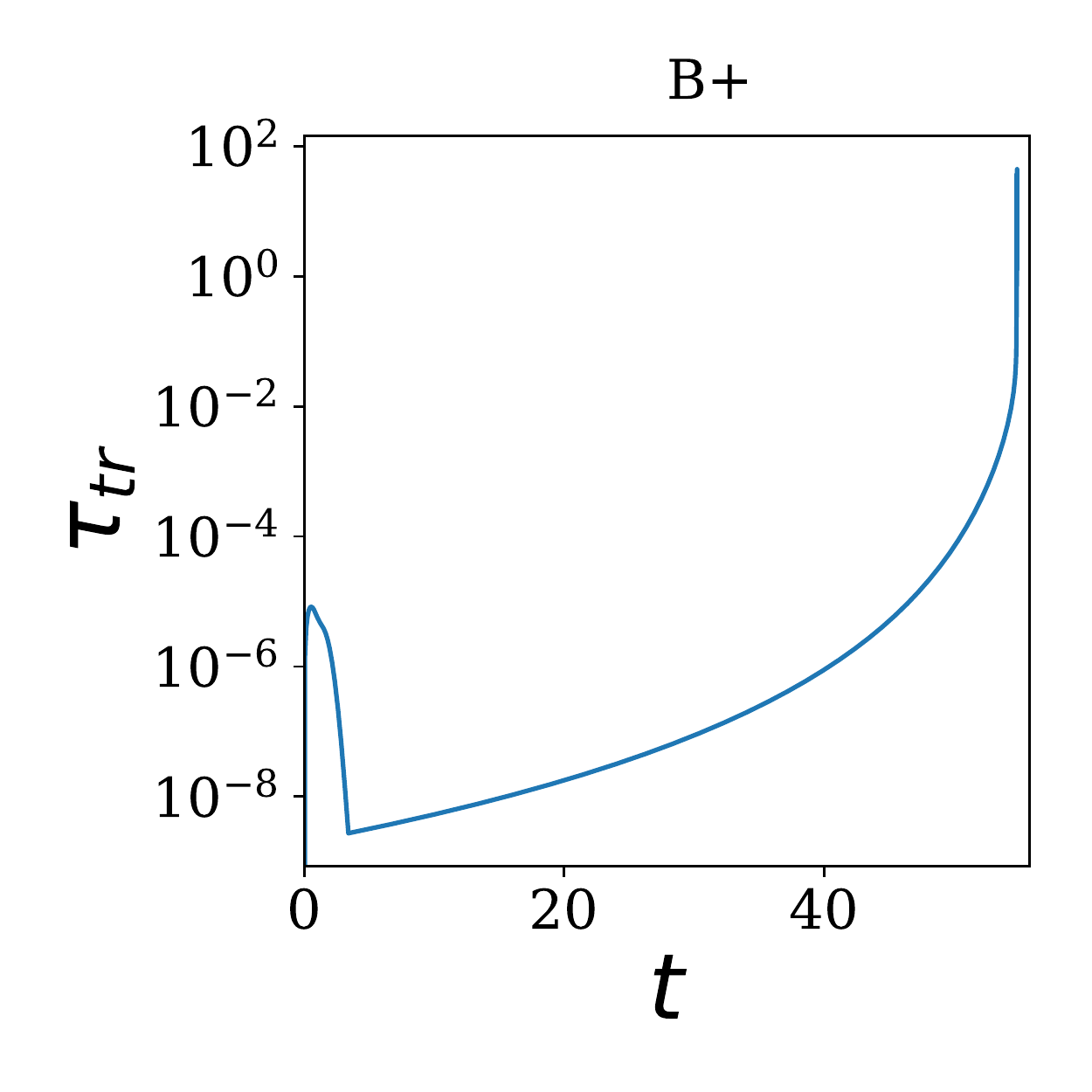}
\includegraphics[width=150pt]{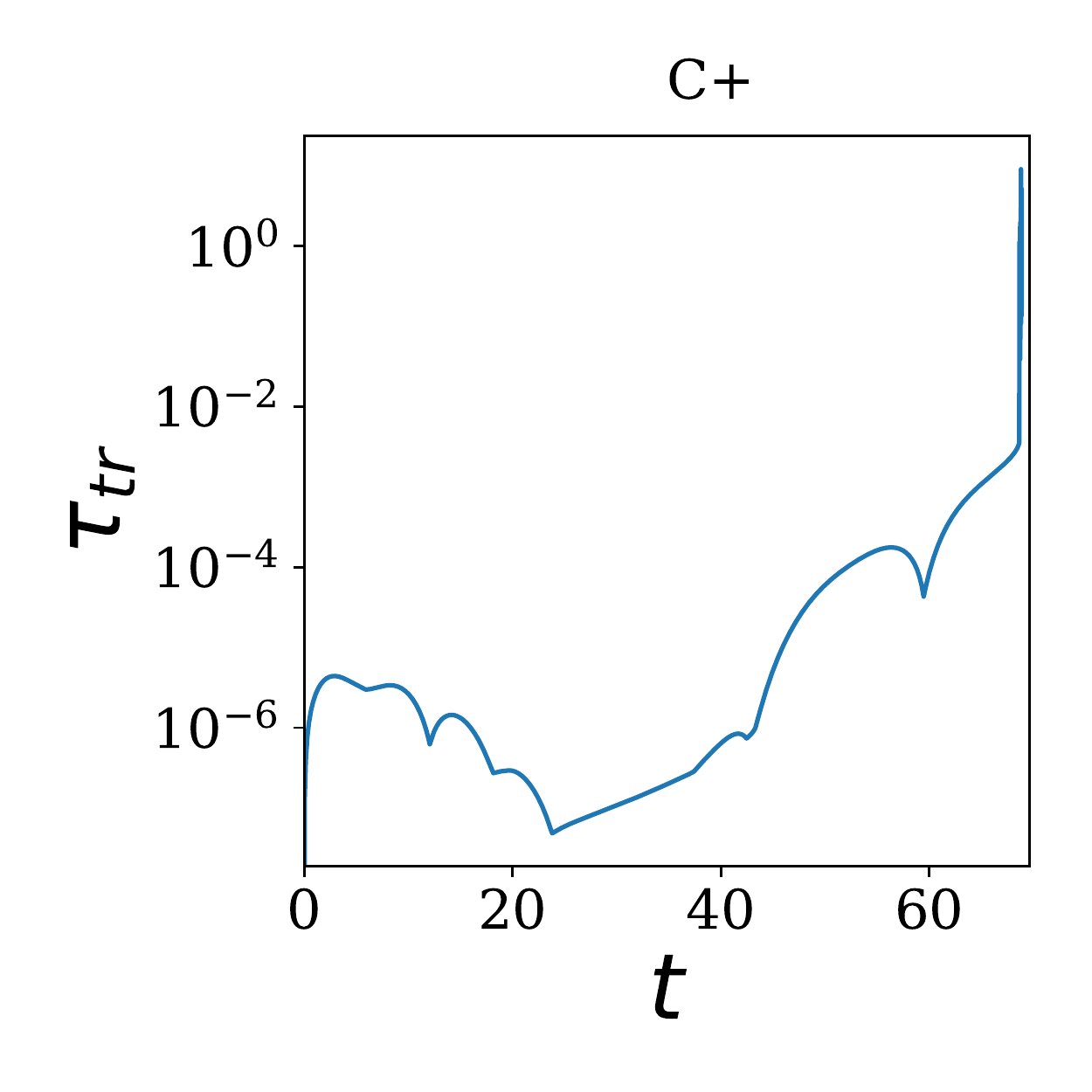}
\caption{$\max|\tau_{tr}|$  for cases A+ (left), B+ (center) and C+ (right).\label{fig:TwistPlus}}
\end{figure}

Finally, we illustrate the behavior of the (only non-trivial) component, $\tau_{tr}$, of the twist 
\begin{equation}
\tau_{\mu \nu} = \nabla_{[\mu}(X\partial_{\nu]}\phi)\, ,
\end{equation}
in figures (\ref{fig:TwistMinus}),(\ref{fig:TwistPlus}) for the negative and positive couplings adopted.
As it is evident in the figures, in the weak cases (A-,A+), the twist remains bounded and relatively 
small throughout the evolution. In contrast, in all but the C- cases  
the twist grows without bound. In case C-, however, the twist remains bounded since the large value of $a$ at the horizon 
causes $X = a^{-2}(\Pi^2 - \Phi^2)/2$ to approach zero.

\section{Final Comments}
In this work we explored the subset of Horndeski's theories identified
as being able to define locally well posed problems. The analysis
we build upon, described in~\cite{Papallo:2017qvl,Ijjas:2018cdm}, relied on 
identifying and exploiting a specific gauge. Such a choice might a priori be regarded
as restrictive, however when seen from the Einstein frame point
of view, it can be argued as being quite natural. Further, note the discussion --and problems
identified that can arise--  for the dynamics of the scalar field holds regardless
of the gauge chosen to consider the evolution for the metric sector. 
In particular, one can argue for the existence of global well behaved solutions in the
weak data case. 
Beyond this regime, however the truly non-linear character of the equations
can induce phenomenology which present serious roadblocks. Avoiding such issues requires 
satisfying a twist-free condition, but such a case 
might be too restrictive depending on the application and context of interest. 
In the general case, the strong possibility of a change in character of the equation --from hyperbolic
to elliptic through a parabolic stage-- as well as the loss of uniqueness through the appearance
of shocks further question the ability to define
well-posed problems with these theories. (In simplified settings, similar deficiencies
have been identified~\cite{Appleby:2011aa,DeFelice:2011bh,Brito:2014ifa}). 
 We mention in passing that since the effective metric depends on the gradient of the scalar
field, the transition to parabolic/elliptic regimes is likely to take prior to the formation of shocks
in generic situations (also highlihgted in~\cite{Ripley:2019hxt}). Hence, considering Horndeski's theories
as the leading order in a gradient expansion, problems might arise still within the a priori assumed regime of applicability.
The timescale for the identified pathologies
to arise depends, naturally, on the coupling value considered and the initial data
adopted.
Due to these difficulties, the extent to which global solutions obtained
within the linearized regime and the information one can draw from them with respect
to the original action can be regarded as suspect.

This observation, which is arguably in tension with interesting observations drawn at
linearized levels in the cosmological context, perhaps calls for a different philosophy
with respect to Horndeski's theories. For instance, to use the linearized equations of
motion as a starting point to build a new one free of the (many) problems identified
at the nonlinear level through the addition of further suitable operators (for a related 
discussion, see~\cite{deRham:2019wjj}). However, it might come at the expense of higher
derivatives being introduced. A complementary or alternative approach would be to identify  the set of behaviors which can be
considered physical and, armed with a suitable justification, modify the non-linear
equations of motion to control unphysical pathologies (e.g.~\cite{Cayuso:2017iqc,Allwright:2018rut}).

\acknowledgements
We are indebted to Gustav Holzegel and Jonathan Luk for discussions on the Null Condition and
the Strauss conjecture. We also thank Anna Ijjas, Frans Pretorius, Oscar Reula and Olivier Sarbach
for discussions.
R.~L. gratefully acknowledges the hospitality of the Perimeter Institute for Theoretical Physics, where this work was done. 
R.~L. is supported by the Spanish Ministerio de Ciencia, Innovaci\'on y Universidades Grant No. FPU15/01414 and by MEC grant FPA2016-76005-C2-2-P.
This research was supported in
part by NSERC, CIFAR and the Perimeter Institute for Theoretical
Physics.  Research at Perimeter Institute is supported by the
Government of Canada through the Department of Innovation, Science and
Economic Development Canada, and by the Province of Ontario through
the Ministry of Research and Innovation.

\appendix

\section{Numerical Details}

In order to have a better time resolution in the last stages of the simulations, the Courant parameter is reduced from 1/10 to 1/100, and the output is produced every 4 time steps. The instants of time where this happens are listed in TABLE \ref{tab:refinement}.

\begin{table}[h!]
\begin{tabular}{|c|c|c|c|c|c|c|}
\hline
Parameter Set & A- & B- & C- & A+ & B+ & C+ \\
\hline 
Refinement Time & Never & $t = 56.5$ & Never & Never & $t = 54.0$ & $t = 68.0$ \\
\hline 
\end{tabular}
\caption{Instants in time where resolution is increased. \label{tab:refinement}}
\end{table}

\begin{figure}[h!]
\includegraphics[width=120pt]{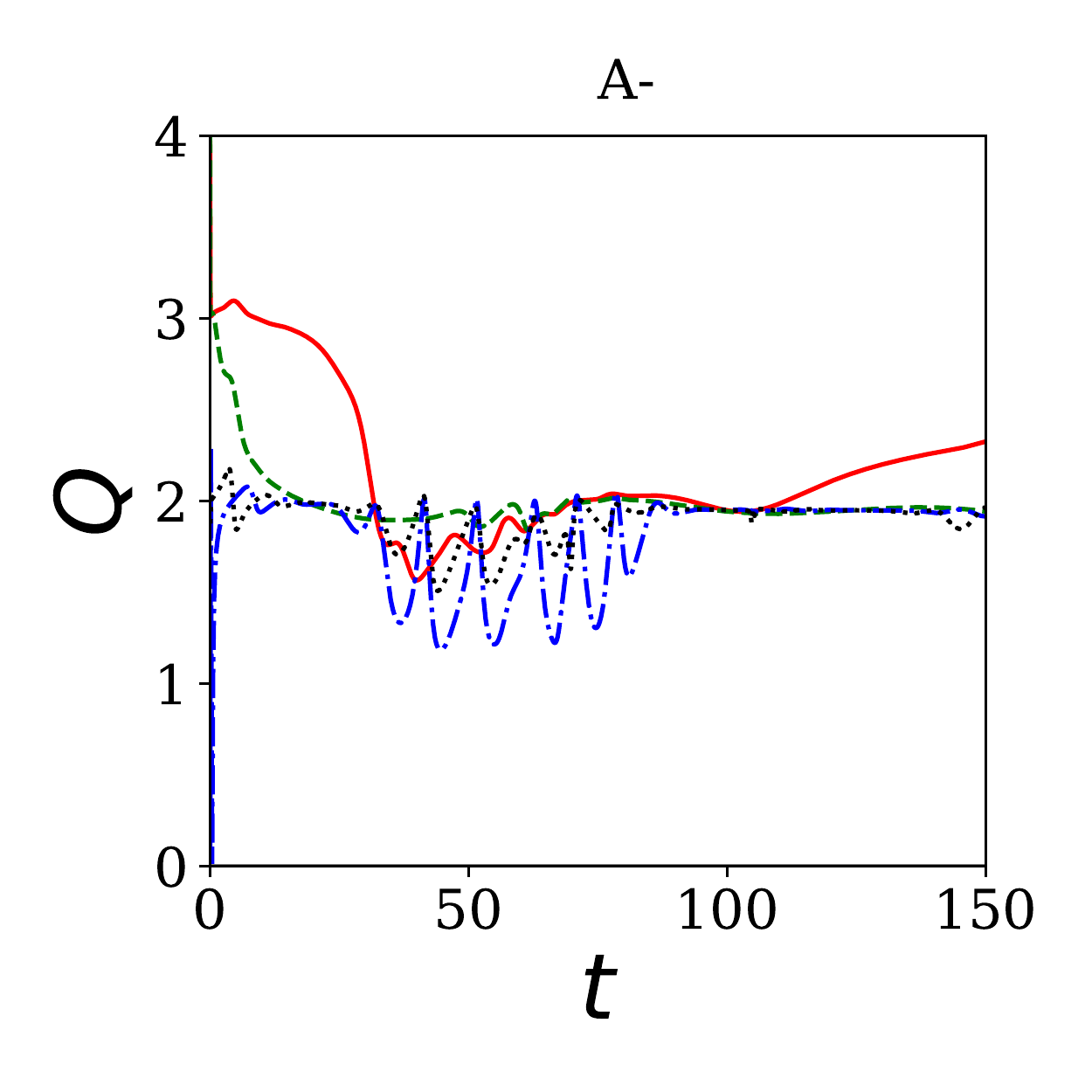}
\includegraphics[width=120pt]{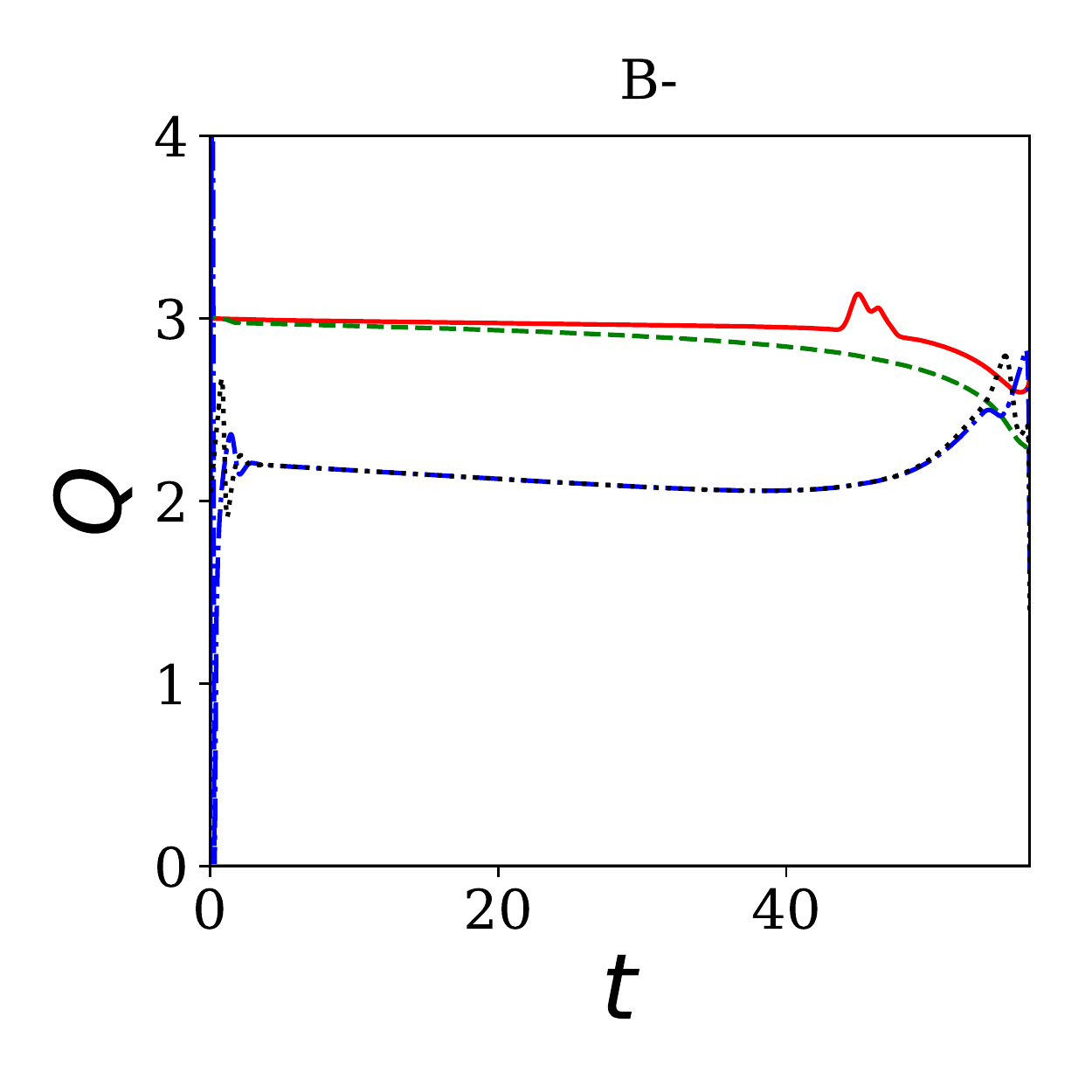}
\includegraphics[width=120pt]{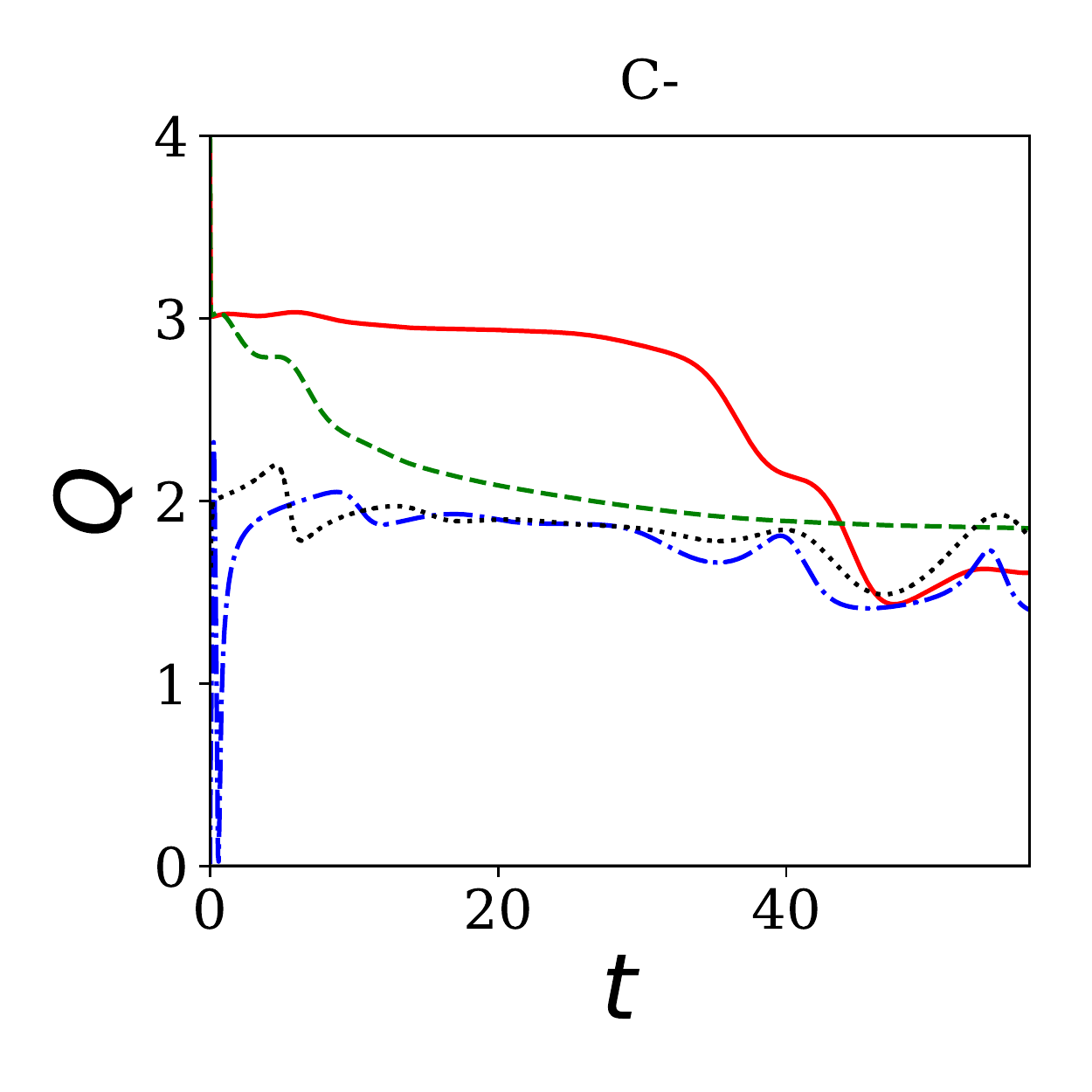}\\
\includegraphics[width=120pt]{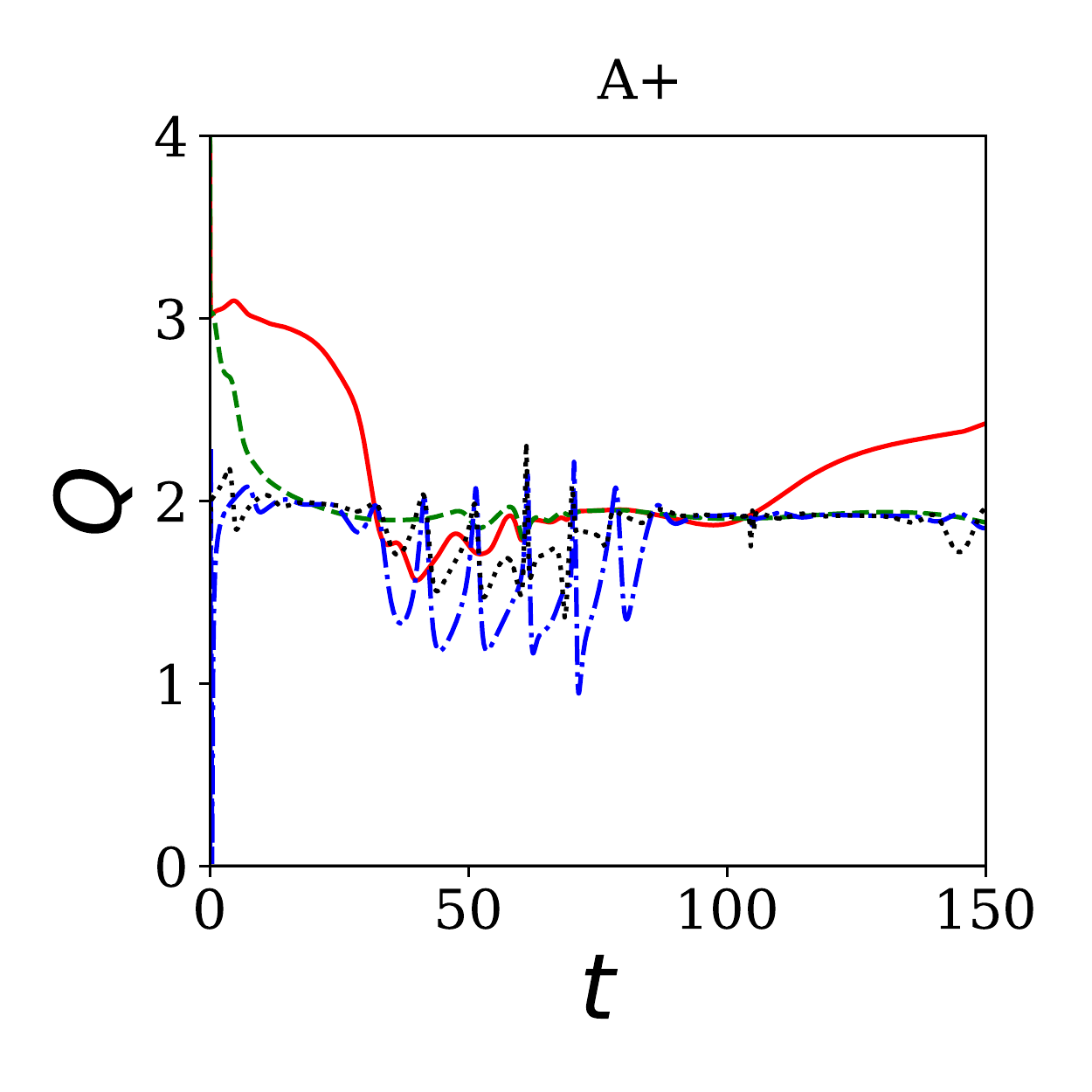}
\includegraphics[width=120pt]{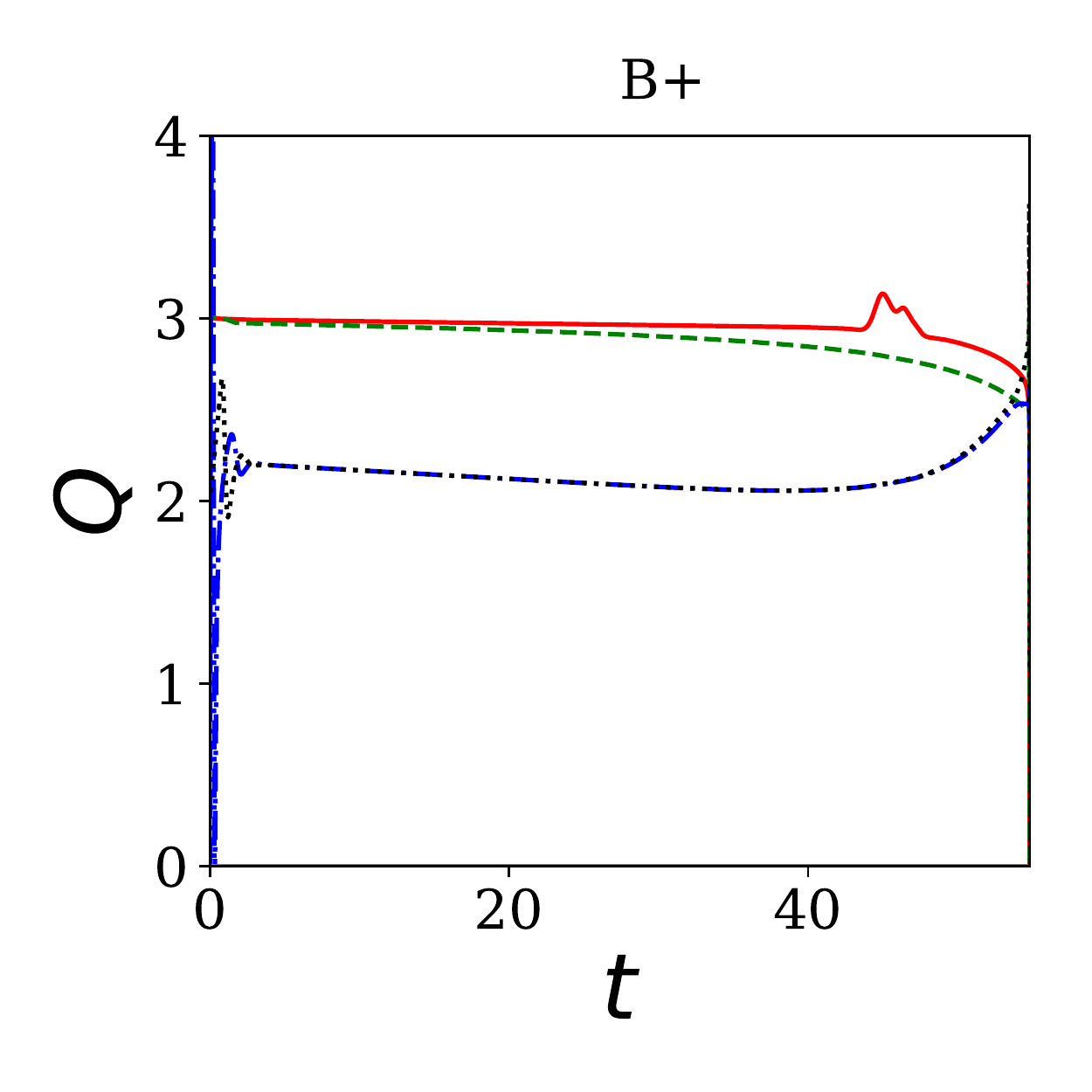}
\includegraphics[width=120pt]{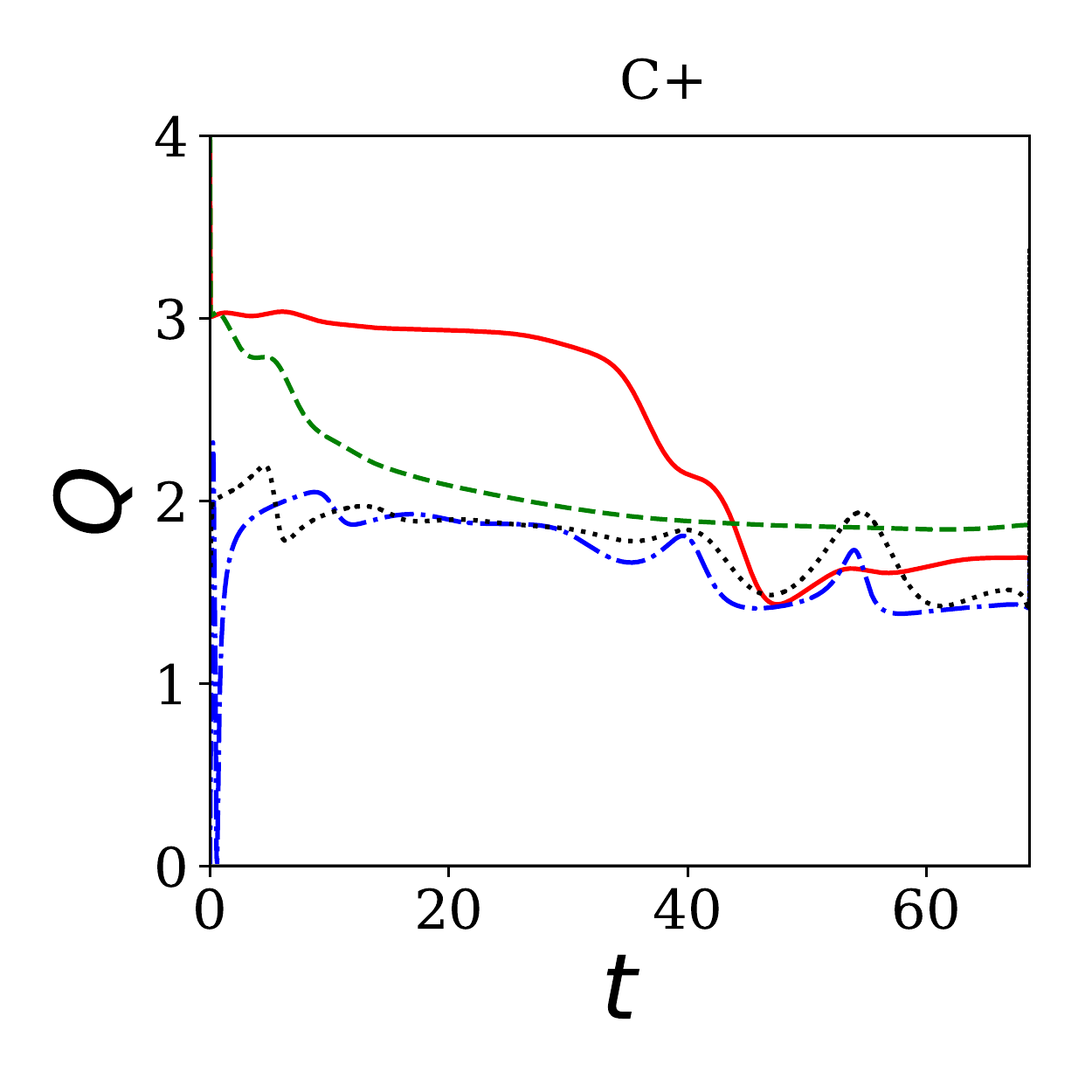}
\caption{Order of convergence $Q$ for the four grid functions $\alpha$ (red solid), $a$ (green dashed), $\Phi$ (Blue, dash-dotted) and $\Pi$ (black dotted).\label{fig:convergence}}
\end{figure}

Further, we compute the order of convergence of solutions $Q$ as

\begin{equation}
 2^Q = \frac{|S_{\Delta /2} - S_{\Delta}|}{|S_{\Delta /4} - S_{\Delta /2}|}\, .
\end{equation}
In FIG. \ref{fig:convergence} the order of convergence is shown as a function of time for the four grid functions,
indicating convergence with the expected rate.

\newpage

\bibliographystyle{ieeetr}
\bibliography{references}

\end{document}